\documentclass[preprint,12pt]{elsarticle}

\usepackage{amssymb}
\usepackage{amsmath}
\usepackage{xcolor}
\usepackage{graphicx}
\usepackage{xr}
\usepackage{csquotes}

\journal{Applied Surface Science}

\begin{document}

\begin{frontmatter}


\title{Tuning Thermal Conductivity and Electron-Phonon Interactions in Carbon and Boron Nitride Moir\'e Diamanes via Twist Angle Manipulation}


\author[inst1]{Rustam Arabov}
\author[inst1,inst2]{Nikita Rybin}
\author[inst3]{Victor Demin}
\author[inst1]{Mikhail Polovinkin}
\author[inst1]{Alexander Kvashnin}
\author[inst3]{Leonid Chernozatonskii}
\author[inst1,inst2]{Alexander Shapeev}

\affiliation[inst1]{organization={Skolkovo Institute of Science and Technology},
addressline={Bolshoy Boulevard, 30, p.1},
city={Moscow},
postcode = {121205},
country={Russia}}

\affiliation[inst2]{organization={Digital Materials LLC},
addressline={Medovaya st., 3},
city={Zarechye, Moscow Region},
postcode = {143085},
country={Russia}}

\affiliation[inst3]{organization={Emanuel Institute of Biochemical Physics RAS},
addressline={Kosygin Street, 4},
city={Moscow},
postcode = {119334},
country={Russia}}

\begin{abstract}

We have investigated the effect of interlayer twist angle on lattice thermal conductivity (LTC) and band gap renormalization in boron nitride and carbon Moir\'e diamanes. Moment tensor potentials were used for calculating energies and forces of interatomic interactions. The methods based on the solution of Boltzmann transport equation (BTE) for phonons and the Green-Kubo (GK) formula were utilized to calculate LTC. The 20-40\% difference in LTC values obtained with GK and BTE-based methods showed the importance of high-order anharmonic contributions to LTC. Significant reduction (by 4.5 - 9 times) of the in-plane LTC with the twist angle increase caused by the growth of structural disorder was observed in the Moir\'e diamanes. This growth of disorder also leads to higher band gap renormalization (induced by classical nuclei motion) in the structures with higher twist angles. Significant band gap renormalization values obtained considering the quantum nuclear effects are caused by the high phonon frequencies related to the bonds with hydrogen atoms on the Moir\'e diamanes surfaces. Understanding of the twist angle effect on LTC and electron-phonon coupling in the Moir\'e diamanes provides a fundamental basis for manipulating their thermal and electronic properties, making these materials promising for thermoelectrics, microelectronics and optoelectronics.

\end{abstract}

\begin{graphicalabstract}
\includegraphics[trim={0cm 0cm 0cm 0cm},clip, width=1\linewidth]{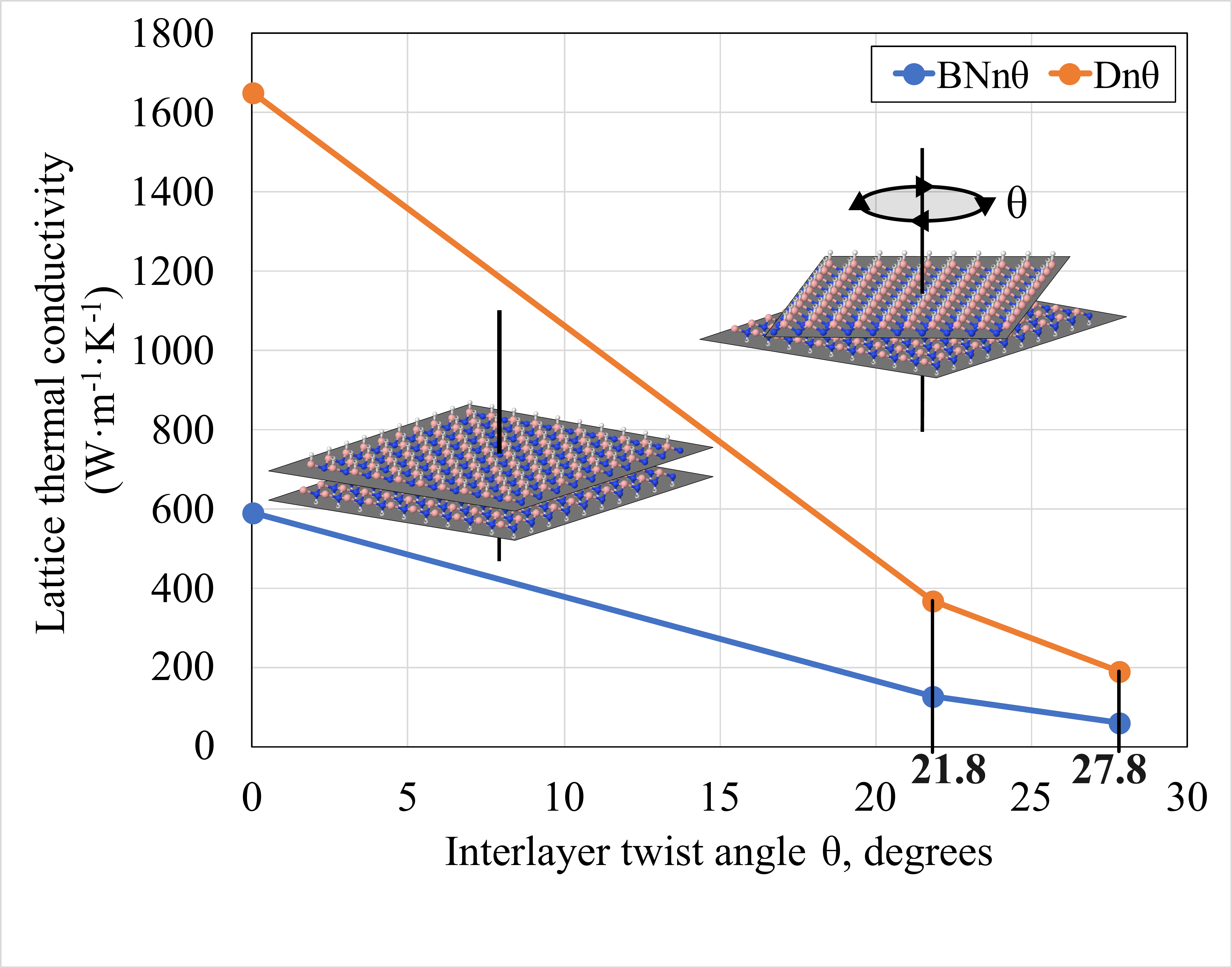}
\end{graphicalabstract}

\begin{highlights}
\item Twist of initial layers of C or BN diamanes increases the structural disorder degree
\item  Increase of disorder reduces lattice thermal conductivity of the Moir\'e diamanes
\item Higher-order anharmonicity affects thermal conductivity in the Moir\'e diamanes
\item Increased structural disorder correlates with higher band gap renormalization values
\item Presence of light atoms on the surface causes high band gap renormalization at T=0 K
\end{highlights}

\begin{keyword}
Moir\'e lattices \sep diamane \sep lattice thermal conductivity \sep band gap renormalization \sep machine learning interatomic potentials \sep boron nitride

\PACS 63.20.kd  \sep 63.22.-m \sep 66.70.Df \sep 34.20.Cf

\MSC 82D37 \sep 82C70 \sep 68T20
\end{keyword}

\end{frontmatter}



\section{Introduction}

Twistronics is a rapidly evolving field of materials science that explores the manipulation of the properties of two-dimensional (2D) layered materials by controlling the angle at which adjacent layers are twisted with respect to each other \cite{twistronics1}.
Twistronics gained prominence after the experimental discovery of superconductivity in twisted bilayer graphene at specific \textquote{magic} angles \cite{cao2018unconventional} -- a phenomenon that had been predicted in theory \cite{twistBGbands}.
Subsequent research has revealed many phenomena in the twisted 2D systems. Mott insulator states \cite{twistronics2}, unconventional superconductivity \cite{twist_prop_2}, unique phonon behaviour features \cite{twistronics4}, and lattice thermal conductivity changes \cite{kvashnin_diamanes, erhart_twist_ltc, moire_graphene} have been observed in several 2D materials at specific interlayer twist angles. 

This work concentrates on the impact of interlayer twisting on electron and phonon properties in Moir\'e diamanes. From a practical point of view, tuning of these properties in 2D materials can be highly beneficial for microelectronics and optoelectronics~\cite{novoselov20162d}. We have chosen to investigate lattice thermal conductivity (LTC) and band gap renormalization (BGR) due to their critical roles in material performance. Both properties are crucial for practical applications of semiconductors in electronic devices. The electronic band gap directly affects electrical conductivity, and efficient thermal management is essential for device performance because electric current generates heat. Furthermore, in semiconducting materials, phonons play a significant role in both thermal and electronic transport. Lattice thermal conductivity dominates thermal transport in semiconductors because the electronic contribution to thermal conductivity is comparatively small. Moreover, electron-phonon coupling causes band gap renormalization, linking electronic transport to phonon behavior. Therefore, understanding both thermal and electronic transport in semiconductors, and their relationship to phonons, is essential for optimizing electronic devices performance. Moreover, accurate prediction of LTC and band gap values necessitates considering anharmonic effects, which play a crucial role in many materials, especially at elevated temperatures \cite{zacharias2020fully, zheng2022anharmonicity}. The importance of anharmonicity is particularly pronounced when investigating BGR and LTC in semiconductors, as discussed in the works \cite{zhu2025effect, seidl2023anharmonic, yuan2024influence}.


In the present work, we examine how the twist angle affects the lattice thermal conductivity and electron-phonon coupling in diamane-like structures such as hydrogenated boron nitride (BNn$\theta$) \cite{B-AL-GA-theoretical} and hydrogenated graphene (Dn$\theta$) \cite{chernozatonskii2021fully} bilayers with a twist angle $\theta$. We compare AB-stacked structures with the twist angle $\theta = 0^\circ$ and Moir\'e lattices with $\theta = 21.8^\circ$ and $27.8^\circ$. LTC was calculated for these structures by means of the approaches based on the solution of Boltzmann transport equation for phonons (BTE-based method) and the Green-Kubo formula (GK method). For studying the electron-phonon coupling in the Moir\'e lattices, the dependencies of electronic band gap renormalization on temperature were obtained. BGR is the difference between the band gaps in the ideal structure at T = 0 K and in the structure at given temperature. In the present work the gaps between adsorption edges were considered as band gaps. Zero-point band gap renormalization (ZPR) is the difference between the band gaps in the structure at T = 0 K obtained with and without considering the quantum effect of nuclei zero-point fluctuations. 

Calculations required for obtaining lattice thermal conductivity and band gap renormalization are computationally expensive using \textit{ab initio} methods such as density functional theory (DFT). To address this issue, we use machine learning interatomic potentials (MLIPs), which provide DFT-level accuracy at significantly reduced computational costs \cite{mlip_proofs_1, mlip_proofs_2, mlip_proofs_3, mlip_proofs_4, mlip_proofs_5}. Specifically, we use the moment tensor potential (MTP) \cite{novikov2020mlip}, which has been shown to accurately capture phonon properties for LTC calculations in many studies \cite{mtp_proofs_1, mtp_proofs_2, mtp_proofs_3, mtp_proofs_4, mtp_proofs_5}.

Comparison of the LTC values for the three BNn$\theta$ and Dn$\theta$ Moir\'e lattices reveals that structures with higher twist angles exhibit lower in-plane LTC at a fixed temperature.
This is due to the reduction in phonon mode lifetimes caused by the growth of structural disorder associated with an increase in the twist angle. The discrepancy between the LTC values obtained using the Green-Kubo and Boltzmann transport equation methods highlights the significant contribution of higher-order anharmonic effects to the LTC in BNn$\theta$ and Dn$\theta$ structures. Additionally, the growth of structural disorder with the increase in the $\theta$ angle leads to higher band gap renormalization values in structures with higher twist angles. Notably, the substantial zero-point renormalization of band gap observed in the Moir\'e diamanes primarily originates from the presence of light hydrogen atoms in the surface layers.

These results demonstrate that functionalization and twist angle manipulation provide an effective approach for modulating both the thermal transport characteristics and the electronic properties in Moir\'e lattices. This opens up new possibilities for designing advanced functional materials with optimized performance for applications in thermoelectrics, optoelectronics, and next-generation microelectronic devices.

\section{Computational methodology}

\subsection{DFT calculations}

DFT calculations of energies and forces of interatomic interactions in crystal lattices were performed using the Vienna Ab initio Simulation Package (VASP)~\cite{VASP}, which utilizes the projector augmented-wave method~\cite{Blochl1994}. All calculations were performed on the basis of the generalized gradient approximation and the Perdew–Burke–Ernzerhof (PBE)~\cite{Perdew1996a} functional, along with Grimme's DFT-D3~\cite{Grimme2010} van der Waals dispersion correction. The plane-wave energy cutoff values were set to 550 eV (for BNn$\theta$) and 500 eV (for Dn$\theta$). 

For the calculations, untwisted AB-stacked bilayers were selected, as well as twisted bilayers with angles of 21.8$^\circ$ and 27.8$^\circ$.
The BNnAB and DnAB were selected as the most energetically favorable diamane-like structure based on the untwisted h-BN \cite{BNH-theoretical} and graphene bilayers. 
The choice of angles is due to the fact that for hexagonal layers there is a set of angles for which periodic structures can be constructed.
The cell of such periodic structures is described by the vectors $A_1$ = m$a_1$+n$a_2$ and $A_2$ = –n$a_1$ + (m + n)$a_2$, where $a_1$ and $a_2$ are the translation vectors of a single graphene or h-BN layers. The smallest possible periodic twist structures have indices (2,1) and (3,1), which correspond to angles of 21.8$^\circ$ and 27.8$^\circ$ ~\cite{BG}.

We used \textit{ab initio} molecular dynamics (AIMD) simulations to collect an initial training set for potential training. In the AIMD simulations, the supercells of sizes $4 \times 4 \times 1$ (96 atoms), $1 \times 1 \times 1$ (46 atoms), and $1 \times 1 \times 1$ (82 atoms) were used for structures with 0$^\circ$, 21.8$^\circ$, and 27.8$^\circ$ twist angles, respectively. 
The corresponding k-point meshes for sampling of the Brillouin zone were set to $2 \times 2 \times 1$, $4 \times 4 \times 1$, and $3 \times 3 \times 1$.

The simulation box sizes along the out-of-plane direction were set to 25 \AA~ (effectively 20 \AA~ of vacuum) for all structures to avoid artificial interaction between periodic images of the 2D structure.
The electronic self-consistent loop treshold was set to 10$^{-6}$ eV for the total energy change and 10$^{-5}$ eV/\AA for the change of Hellman–Feynman forces.
During structural relaxation, the conjugate gradient algorithm was used until the forces dropped below 10$^{-5}$ eV/\AA.
This was done to exclude interactions along the out-of-plane direction between the layers from the neighboring simulation boxes in the three-dimensional periodic systems.

\subsection{Moment Tensor Potential}

In this work, Moment Tensor Potentials (MTPs) implemented in the MLIP-2 software package \cite{novikov2020mlip} were used. MTP is a local potential, i.e. the energy of an atomic configuration is a sum of the energies of local atomic environments of the individual atoms:
    \begin{equation}
        E^{\text{MTP}} = \sum_{i=1}^{N} V({\mathfrak{\boldsymbol n}}_{i}),
    \end{equation}
    where the index $i$ labels $N$ atoms of the system, ${\mathfrak{\boldsymbol n}}_{i}$ denotes the local atomic neighborhood around atom \textit{i} within a certain cut-off radius $R_\text{cut}$, and the function $V$ is the energy of atomic neighborhood: 
    \begin{equation}
        V({\mathfrak{\boldsymbol n}}_{i}) = \sum_{\alpha} \xi_{\alpha} B_{\alpha}({\mathfrak{\boldsymbol n}}_{i}).
    \end{equation}
    where $\xi_\alpha$ are the linear parameters to be fitted and $B_\alpha$ are the basis functions that will be defined below. As fundamental symmetry requirements, all descriptors for atomic environment have to be invariant to translation, rotation, and permutation with respect to the atomic indexing. Moment tensors descriptors $M_{\mu, \nu}$ satisfy these requirements and are used as representations of atomic environments: 
    \begin{equation}
        M_{\mu, \nu}\left({\mathfrak{\boldsymbol n}}_{i}\right)=\sum_j f_\mu\left(\left|r_{i j}\right|, z_i, z_j\right) \underbrace{{\boldsymbol r}_{i j} \otimes \ldots \otimes \boldsymbol{r}_{i j}}_{\nu \text { times }},
    \end{equation}
    where the index $j$ goes through all the neighbors of atom $i$. The symbol ``$\otimes$'' indicates the outer product of vectors, thus ${\boldsymbol r}_{i j} \otimes \cdots \otimes {\boldsymbol r}_{i j}$ is the tensor of rank $\nu$ encoding the angular part. 
    The function $f_\mu$ represents the radial component of the moment tensor:
    \begin{equation}
        f_\mu\left(\left|{\boldsymbol r}_{i j}\right|, z_i, z_j\right)=\sum_{\beta} c_{\mu, z_i, z_j}^{(\beta)} Q^{(\beta)}(|{\boldsymbol r}_{i j}|),
    \end{equation}
    where $z_i$ and $z_j$ denote the atomic species of atoms $i$ and $j$, respectively, ${\boldsymbol r}_{ij}$ describes the positioning of atom $j$ relative to atom $i$, $c_{\mu, z_i, z_j}^{(\beta)}$ are the radial parameters to be fitted and 
    \begin{equation}
        Q^{(\beta)}(|{\boldsymbol r}_{i j}|)=T^{(\beta)}(|{\boldsymbol r}_{i j}|)\left(R_{\text {cut }}-|{\boldsymbol r}_{i j}|\right)^2
    \end{equation}
    are the radial functions consisting of the Chebyshev polynomials $T^{(\beta)}(\left|{\boldsymbol r}_{i j}\right|)$ on the interval $(R_\text{min},  R_\text{cut})$. The number of the radial parameters scales quadratically with the number of atomic species in structures. The descriptors $M_{\mu, \nu}$ with $\nu$ equal to $0, 1, 2, \ldots$ are tensors of different ranks that allow one to define basis functions as all possible contractions of these tensors to a scalar. It is possible to construct an infinite number of $B_{\alpha}$ and in order to restrict this number in the MTP functional form, the level of moment tensor descriptors ${\rm lev}M_{\mu, \nu}$ = 2 + 4$\mu$ + $\nu$ was introduced. If $B_{\alpha}$ is obtained from $M_{\mu_1, \nu_1}$, $M_{\mu_2, \nu_2}$ ,
    $\dots$, then ${\rm lev}B_{\alpha}$ = $(2 + 4\mu_1 + \nu_1)$ + $(2 + 4\mu_2 + \nu_2)$ + $\dots$. Therefore, MTP of level $d$ can be obtained by including a finite number of basis functions with ${\rm lev}B_{\alpha} \leq d$. The total set of parameters to be found is denoted by ${\boldsymbol \theta} = (\{ \xi_{\alpha} \}, \{ c^{(\beta)}_{\mu, z_i, z_j} \})$ and the MTP energy of a structure is denoted by $E^{\text{MTP}} = E({\boldsymbol \theta})$.

    \subsection{Extrapolation grade of structures and active learning}\label{extrapolation_grade}
    
    A feature of the MLIP-2 software package is the ability to perform both training of potentials on independently selected configurations and active learning with automatic pre-selection of new configurations. In the second case, in order to constrcut a training set for MTP fitting, it is necessary to calculate the so-called extrapolation grade of structures. For estimating this grade the following steps should be performed. Assume that the initial training set contains $K$ structures with energies $E^{\rm DFT}_k$, forces ${\boldsymbol f}^{\rm DFT}_{i,k}$, and stresses $\sigma^{\rm DFT}_{i,k}$, $k = 1, \ldots, K$ calculated within DFT framework. First, the initial MTP is trained, i.e. by solving the following minimization problem, optimal parameters ${\boldsymbol{\bar{\theta}}}$ are found:
    \begin{equation}
        \begin{array}{c}
            \displaystyle
            \sum \limits_{k=1}^K \Bigl[ w_{\rm e} \left(E^{\rm DFT}_k - E_k({\boldsymbol {\theta}}) \right)^2 + w_{\rm f} \sum_{i=1}^N \left| {\boldsymbol f}^{\rm DFT}_{i,k} - {\boldsymbol f}_{i,k}({\boldsymbol {\theta}}) \right|^2 
            \\
            \displaystyle
            + w_{\rm s} \sum_{i=1}^6 \left| \sigma^{\rm DFT}_{i,k} - \sigma_{i,k}({\boldsymbol {\theta}}) \right|^2 \Bigr] \to \operatorname{min},
        \end{array}
    \end{equation}
    where $w_{\rm e}$, $w_{\rm f}$, and $w_{\rm s}$ are non-negative weights expressing the importance of energies, forces, and stresses in the minimization problem. 
    
    After finding the optimal parameters of ${\boldsymbol{\bar{\theta}}}$, it is necessary to create the following matrix:
    \begin{equation}
        \mathsf{B}=\left(\begin{matrix}
            \frac{\partial E_1\left( {\boldsymbol{\bar{\theta}}} \right)}{\partial \theta_1} & \ldots & \frac{\partial E_1\left( {\boldsymbol{\bar{\theta}}} \right)}{\partial \theta_m} \\
            \vdots & \ddots & \vdots \\
            \frac{\partial E_K\left( {\boldsymbol{\bar{\theta}}} \right)}{\partial \theta_1} & \ldots & \frac{\partial E_K\left( {\boldsymbol{\bar{\theta}}} \right)}{\partial \theta_m} \\
        \end{matrix}\right),
    \end{equation}
    where each row corresponds to a particular structure. The matrix $\mathsf{B}$ is used to construct a subset of structures yielding the most linearly independent rows (physically it means geometrically different structures), which is equivalent to finding a square $m \times m$ submatrix $\mathsf{A}$ of the matrix $\mathsf{B}$ of maximum volume (maximal value of $|{\rm det(\mathsf{A})}|$). To achieve this, the so-called maxvol algorithm is applied \cite{goreinov2010maxvol}. To determine whether a given structure $\boldsymbol x^*$ obtained during an atomistic simulation is representative, the extrapolation grade $\gamma(\boldsymbol x^*)$ should be calculated, which is defined as
    \begin{equation} \label{Grade}
        \begin{array}{c}
            \displaystyle
            \gamma(\boldsymbol x^*) = \max_{1 \leq j \leq m} (|c_j|), ~\rm{where}
            \\
            \displaystyle
            {\boldsymbol c} = \left( \dfrac{\partial E}{\partial \theta_1} (\boldsymbol{\bar{\theta}}, \boldsymbol x^*) \ldots \dfrac{\partial E}{\partial \theta_m} (\boldsymbol{\bar{\theta}}, \boldsymbol x^*) \right) \mathsf{A}^{-1}.
        \end{array}
    \end{equation}
    This grade specifies the maximal factor by which the determinant $|{\rm det(\mathsf{A})}|$ can increase if ${\boldsymbol x^*}$ is added to the training set. Thus, if the structure $\boldsymbol x^*$ is a candidate for adding to the training set then $\gamma( x^*) \geq \gamma_{\rm th}$, where $\gamma_{\rm th} \geq 1$ is an adjustable threshold parameter which controls the value of permissible extrapolation. Otherwise, the structure is not representative and hence it is not included into the training set.

\subsection{Moment tensor potential training}

The initial training sets were created from the AIMD trajectories for each Moir\'e lattice.
The AIMD simulations were performed with the canonical (NVT) ensemble using the Nos\'e-Hoover thermostat \cite{nose-hoover} at a constant temperature of 1500 K.
The total time of each AIMD simulation was 1 ps with a timestep of 1 fs.
The resulting trajectories were subsampled at 5 fs time intervals.
The resulting sets of 200 snapshots for each structure were employed to train initial MTPs of level 10 (for each Moir\'e lattice, the MTPs were trained separately).
After that, the active learning of MTP was performed with the algorithm implemented in the MLIP-2 package \cite{novikov2020mlip}.
This algorithm works as follows. 
Firstly, MD simulations in the LAMMPS package \cite{lammps} with MTP as a model for interatomic interaction are performed in the NVT ensemble (at a temperature of 1500 K for the structures with zero and 21.8$^\circ$ twist angles). The temperature of MD simulation for the structures with 27.8$^\circ$ twist angle was decreased to 1000 K because these structures possessed instability at T = 1500 K.
At each step of the MD simulations, the algorithm calculates the extrapolation grade $\gamma$ for the current atomic configuration. Configurations with $\gamma$ higher than 2.1 are added to the preselected set. When $\gamma$ exceeds 10, the MD simulation stops and all sufficiently different configurations from the preselected set are added to the training set. After that, the MTP is refitted. This procedure repeats until the MD simulation runs without failure (i.e., without stops due to $\gamma$ values higher than 10) for 30 ps.
The training set generated during the active learning procedure was used for training the MTP of level 12. Subsequent active learning of the MTP of level 12 led to the addition of new configurations to the training set. This enlarged training set was finally used for fitting the MTP of level 16. The MTP of level 16 was also refined through active learning and subsequently used for lattice dynamics calculations. The numbers of parameters in MTPs of different levels used for studying lattice dynamics in Dn$\theta$ and BNn$\theta$ structures are given in Table S1 in the Supplementary information.

\subsection{Lattice dynamics calculations}

The accuracy of the MTP was validated by assessing the agreement between MTP predictions and DFT results. This agreement was checked only for the structures without interlayer rotation due to high computational costs of DFT-based phonon properties calculations for the twisted structures (the numbers of required calculations are given in the Supplementary Information, Table S2).

The MTP validation procedure was the following. Firstly, the MTP-predicted forces were compared with the values from the testing dataset consisting of 1000 samples, which was generated by performing additional AIMD simulations for 1 ps with a 1 fs time step at a temperature of 1000 K (for the DnAB diamane) and 500 K (for other Moir\'e lattices). Secondly, the phonon properties important for LTC calculations (phonon band structures, heat capacities and group velocities) obtained from DFT and MTP forces were compared. The calculations of the phonon properties were performed in the harmonic approximation, as implemented in the Phonopy package \cite{Togo2023}.

After the MTP validation, lattice thermal conductivity calculations were performed for the BNn$\theta$ and Dn$\theta$ structures. The first method utilized for obtaining LTC in the present work was based on solving the Boltzmann transport equation in the relaxation time approximation, as implemented in the Phono3py package~\cite{Togo2015_phono3py, Togo2018, Togo2023}. The BTE-based method required calculations of the second- and third-order force constants. These calculations were performed using the supercell approach ~\cite{Parlinski1997} with the atomic displacements of 0.03 \AA. While obtaining the force constants for each structure, interactions between all atoms in the supercell were considered. The second-order force constants were used for obtaining the phonon dispersion relations. Then, the heat capacities and group velocities of phonons were calculated from the dispersion relations using the formulas (\ref{eq:heat_cap}) and (\ref{eq:gv}), respectively \cite{Togo2015_phonons}.

\begin{equation}
   C_{q\nu} = k_\mathrm{B}
\left(\frac{\hbar\omega_{q\nu}}{k_\mathrm{B} T} \right)^2
\frac{\exp(\hbar\omega_{q\nu}/k_\mathrm{B}
T)}{[\exp(\hbar\omega_{q\nu}/k_\mathrm{B} T)-1]^2}
\label{eq:heat_cap}
\end{equation}

\begin{equation}
    \textbf{v}_{q\nu} = \frac{\partial\omega_{q\nu}}{\partial q}
    \label{eq:gv}
\end{equation}
where $C_{q \nu}$, $v_{q \nu}$ and $\omega_{q \nu}$ denote the heat capacity, group velocity and frequency of the phonon mode $\nu$ at the point $q$ of Brillouin zone, respectively. 

As a final step of the BTE-based approach, LTC was calculated using the following formula:

\begin{equation}
    \kappa_L = \frac{1}{NV_c}\sum_{q\nu}\tau_{q\nu} C_{q\nu}\textbf{v}_{q\nu}\otimes\textbf{v}_{q\nu}
    \label{eq:kappa}
\end{equation}
where $C_{q \nu}$, $v_{q \nu}$ and $\tau_{q \nu}$ denote the heat capacity, group velocity and lifetime of the phonon mode $\nu$ at the point $q$ of Brillouin zone, respectively, all of the present phonon modes are included in the summation.
$V_c$ is the unit cell volume, $N$ is the number of unit cells in the supercell. 

The phonon lifetimes were calculated from the imaginary part of the phonon self-energies. For this, the third-order force constants were computed \cite{Togo2015_phono3py, Togo2023}.
For the Moir\'e lattices with non-zero twist angles, the computation of phonon lifetimes is resource-consuming due to the low symmetry of these structures and large numbers of atoms in the unit cells. For this reason the calculations were parallelized over both phonon bands and grid points. 

The BTE-based LTC calculation approach implemented in Phono3py employs a third-order Taylor expansion of the material potential energy surface (PES). Consequently, the model only accounts for three-phonon scattering processes,  neglecting the influence of four-phonon and higher-order interactions of phonons. This approach is invalid for highly anharmonic materials, where these higher-order terms are essential for accurate LTC predictions \cite{LTC_BNC_2D, LTC_anharm_importance_1, LTC_anharm_importance_2}.

In the case of highly anharmonic materials, the method based on the Green-Kubo formula can be applied. This approach allows taking into account all orders of the PES anharmonicity during the LTC calculations \cite{baroni, carbogno, knoop, castellano}. In the Green-Kubo method \cite{lammps_GK}, the LTC is determined by integrating the heat current autocorrelation function (HCACF) using the following formula:
\begin{equation}
    \kappa_\alpha = \frac{1}{3k_BT^2V}\int_0^\infty\langle J_\alpha(t) J_\alpha(0)\rangle dt
\end{equation}
where $t$ is the time, $V$ and $T$ are the system volume and temperature,  $J_\alpha$ is the $\alpha$ component of the lattice heat current vector $J$, $\alpha$ is the Cartesian index. $\langle J_\alpha(t) J_\alpha(0)\rangle$ is the heat current autocorrelation function averaged over the ensemble.

To compute the heat flux required for the HCACF, molecular dynamics simulations with the MTP potential (MTP-MD) were performed using the LAMMPS software package \cite{lammps} and the LAMMPS-MLIP interface \cite{mlip_lammps_GK}. The MTP-MD trajectories were divided into segments of 250 ps and 350 ps duration for the DNnAB and BNnAB, respectively. For the other Moir\'e lattices, the duration of each segment was 100 ps. The HCACF was calculated for each segment and then averaged across all segments. The details on how the segments durations were chosen are described in Supplementaty Infomation (see "Description of the correlation time and supercell size choice for LTC calculations in the Green-Kubo approach" section).  Subsequently, the LTC was obtained by integrating the averaged HCACF. Each MTP-MD trajectory lasted for a total of 4 ns (for the BNnAB) and 3.5 ns (for the other diamanes), comprising a 0.5 ns equilibration period in the NVT ensemble followed by a production run in the NVE ensemble, using a time step of 1 fs. The MTP-MD simulations employed rectangular supercells containing the following numbers of atoms: 2400 (0$^\circ$ twist angle), 2944 (21.8$^\circ$ twist angle), and 2952 (27.8$^\circ$ twist angle). The details on how the supercell sizes were selected are also given in Supplementary Information, "Description of the correlation time and supercell size choice for LTC calculations in the Green-Kubo approach" section. Final LTC values for each structure were determined by averaging the results obtained from 8 independent MTP-MD trajectories.

Finally, the results of LTC calculations were compared for the structures with different twist angles in order to understand how the interlayer twisting affects the thermal transport in Moir\'e lattices.

\subsection{Band gap renormalization}

After investigating the impact of twist angle on the LTC, it is equally important to consider the influence of the interlayer twisting on the Moir\'e lattices electronic properties. Atomic vibrations, which govern phonon transport and LTC, also play a crucial role in renormalization of the electronic band gap via electron-phonon coupling. For studying the effect of twist angle on electron-phonon coupling, the dependencies of BGR on temperature were obtained for each Moir\'e lattice. The procedure of BGR calculation was the following. Firstly, the supercells with distorted atoms were generated. In the case of classical nuclei, NVT molecular dynamics simulations lasting for 150 ps with a 0.5 fs timestep were conducted. For each temperature value, 100 samples separated by 0.5 ps from each other were taken from the corresponding MD trajectory. For studying the effect of lattice thermal expansion (LTE) on BGR, NpT molecular dynamics simulations were carried out with a 1 fs timestep over a duration of 150 ps. Pressure was set to 0 during these simulations. For each temperature value, 100 samples separated by 0.5 ps from each other were taken from the corresponding MD trajectory. All these molecular dynamics simulations were performed by means of LAMMPS software \cite{lammps}. To study the influence of nuclear quantum effects on band gap renormalization, other rattled structures were generated by means of quantum harmonic sampling, as implemented in the HIPHIVE package \cite{hiphive}. The distances and directions for shifting the atoms in this approach were obtained in the harmonic approximation using quantum statistics to describe the nuclear motion. Secondly, for each of the atomic configurations considered, a single-point calculation in the FHI-aims software \cite{fhiaims} was performed to obtain the band gap. The PBE exchange-correlation functional \cite{Perdew1996a} was used. All elements were treated with \textit{light} default settings for the basis and integration grids. The k-point sampling was performed using a $2\times2\times1$ Monkhorst-Pack grid. The band gaps were estimated at temperatures of 0 K, 100 K, 300 K, 500 K, and 700 K. At each temperature, the band gap was calculated by averaging over 100 uncorrelated samples.

Finally, the dependencies of BGR on temperature were compared for the structures with different twist angles in order to understand how the interlayer twisting affects the electron-phonon coupling in the Moir\'e lattices. In addition, the values of zero-point band gap renormalization in the Moir\'e lattices were compared with the ZPR values given in the work \cite{gonze_bgr_database} for other materials to estimate if these values were high or small. 

\section{Results and discussion}

\begin{figure}[h!]
	\begin{minipage}[h]{0.8\linewidth}
		\centering{\includegraphics[trim={0cm 0cm 0cm 0cm},clip, width=1\linewidth]{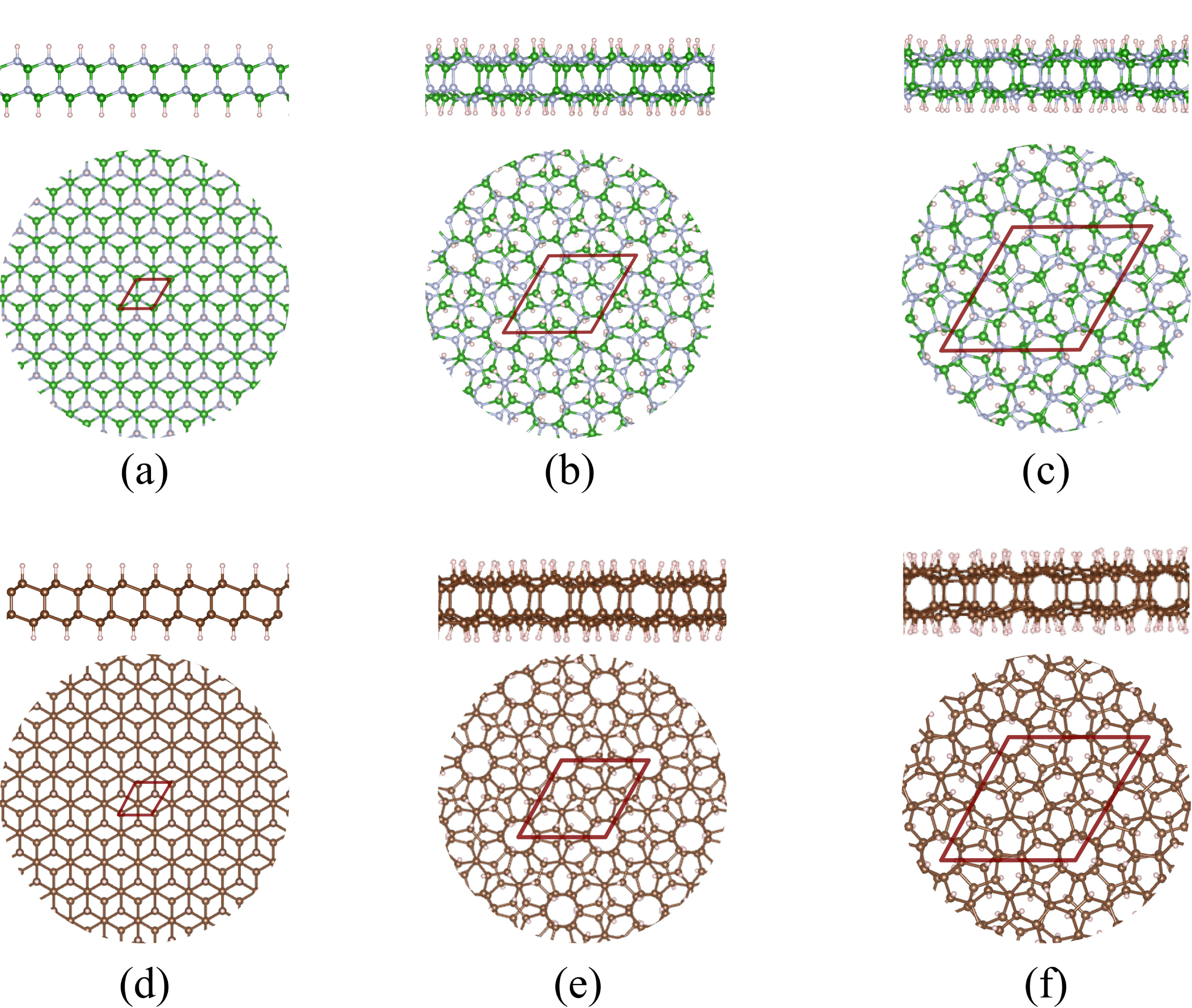}}
	\end{minipage}
  
	\caption{Top and side views of the atomic structure of considered hydrogenated BN and graphene bilayers with twist: BNnAB (a), BNn21.8 (b), BNn27.8 (c), DnAB (d), Dn21.8 (e), Dn27.8 (f) obtained after relaxation with moment tensor potential (MTP).} 

	\label{fig:bnh-ch-scheme}
\end{figure}

The lattice parameters of the Moir\'e lattices obtained after relaxation with the actively trained MTP are given in Tables \ref{table:bnh-params} and \ref{table:ch-params}, and the atomic structures of the Moir\'e lattices are shown in Fig. \ref{fig:bnh-ch-scheme}.  

\begin{table}[h]
\caption{Unit cell parameters of BNn$\theta$ structures after relaxation with MTP.}
\centering
\begin{tabular}{|c|c|c|c|c|c|c|}
\hline
  Structure & $a(=b)$, \AA & $c$, \AA & $\alpha$ & $\beta$ & $\gamma$ & $N_{atoms}$ \\ 
  \hline
  BNnAB & 2.55 &  24.87 & 90.00$^\circ$ & 90.00$^\circ$ & 60.00$^\circ$ & 6\\
  \hline
BNn21.8 & 6.55 &  25.44 & 89.99$^\circ$ & 90.00$^\circ$ & 120.00$^\circ$ & 46\\
  \hline
  BNn27.8 & 9.18 &  25.43 & 91.79$^\circ$ & 90.07$^\circ$ & 120.00$^\circ$ & 82\\
  \hline

\end{tabular}
\label{table:bnh-params}
\end{table}

\begin{table}[h]
\caption{Unit cell parameters of Dn$\theta$ structures after relaxation with MTP.}
\centering
\begin{tabular}{|c|c|c|c|c|c|c|}
\hline
  Structure & $a(=b)$, \AA & $c$, \AA & $\alpha$ & $\beta$ & $\gamma$ & $N_{atoms}$ \\ 
  \hline
  DnAB & 2.58 &  24.84 & 90.00$^\circ$ & 90.00$^\circ$ & 60.00$^\circ$ & 6\\
  \hline
Dn21.8 & 6.72 &  24.69 & 89.99$^\circ$ & 90.00$^\circ$ & 120.00$^\circ$ & 46\\
  \hline
  Dn27.8 & 9.06 &  24.76 & 91.82$^\circ$ & 90.07$^\circ$ & 120.00$^\circ$ & 82\\
  \hline

\end{tabular}
\label{table:ch-params}
\end{table}

\subsection{MTP validation}

\begin{figure*}[h!]
	\begin{minipage}[h]{0.32\linewidth}
		\centering{\includegraphics[trim={0cm 0cm 0cm 0cm},clip, width=1\linewidth]{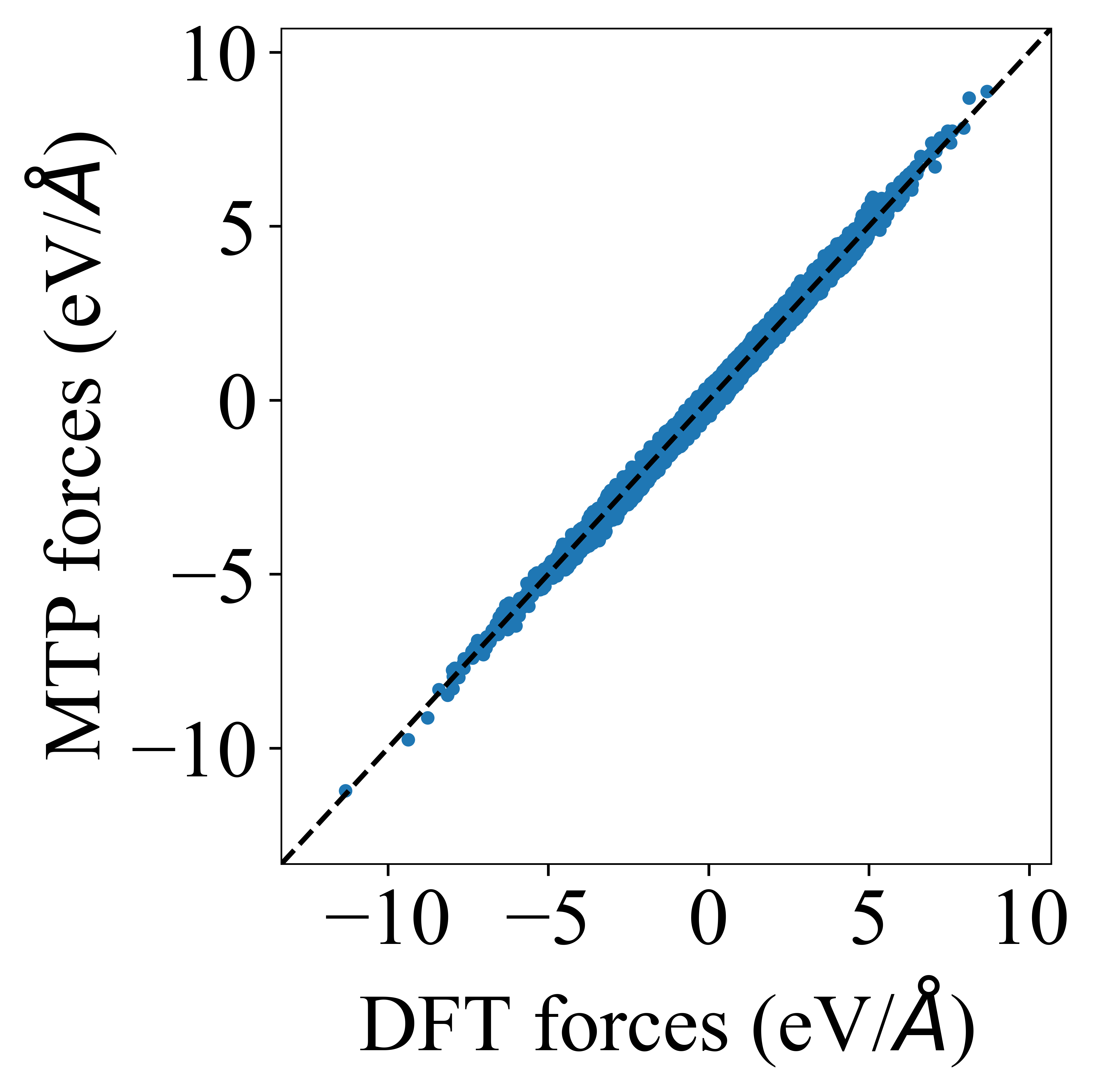}}
        (a)
	\end{minipage}%
    \begin{minipage}[h]{0.32\linewidth}
            \centering{\includegraphics[trim={0cm 0cm 0cm 0cm},clip, width=1\linewidth]{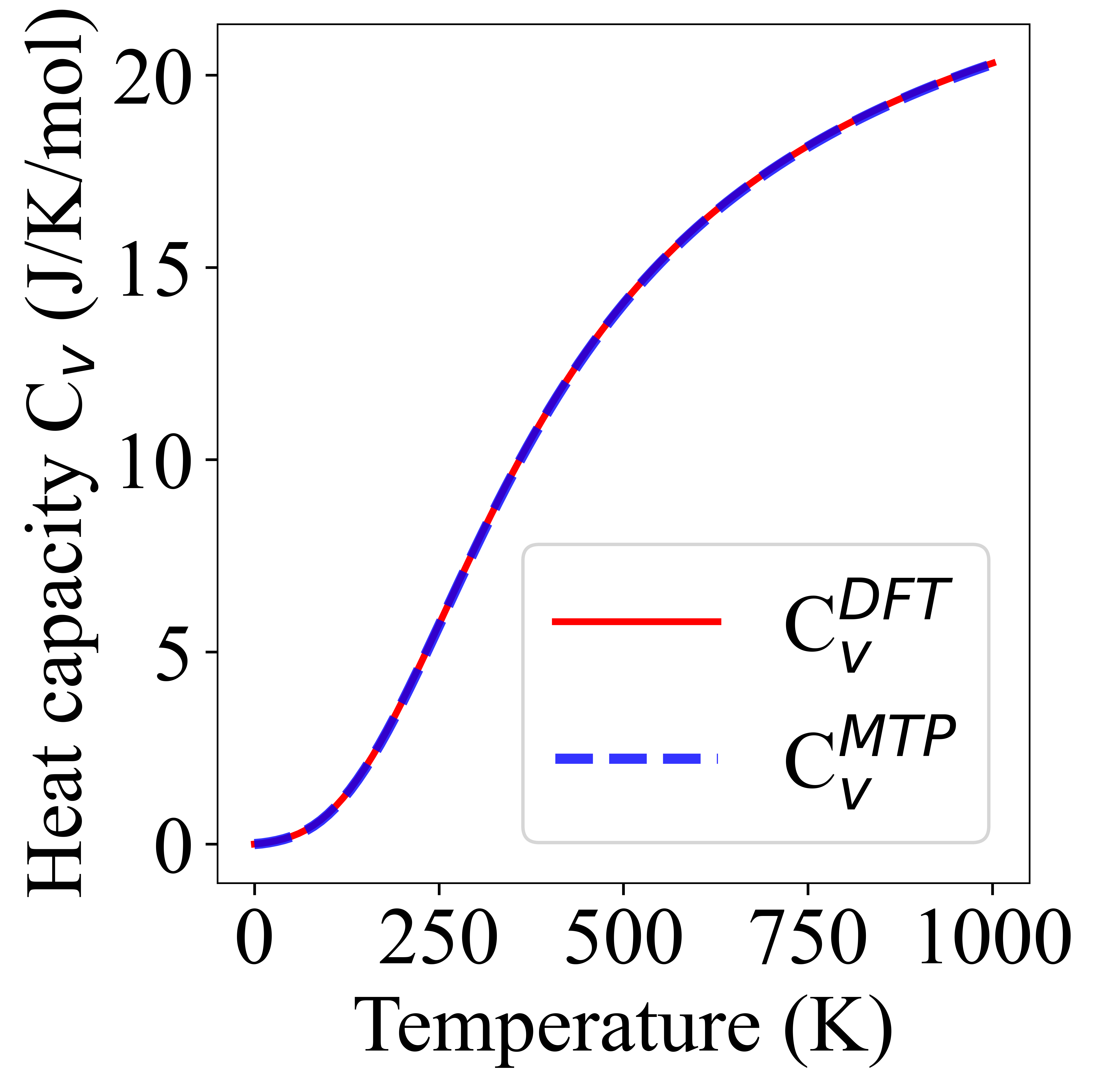}}
        (b)
	\end{minipage}%
    \begin{minipage}[h]{0.32\linewidth}
            \centering{\includegraphics[trim={0cm 0cm 0cm 0cm},clip, width=1\linewidth]{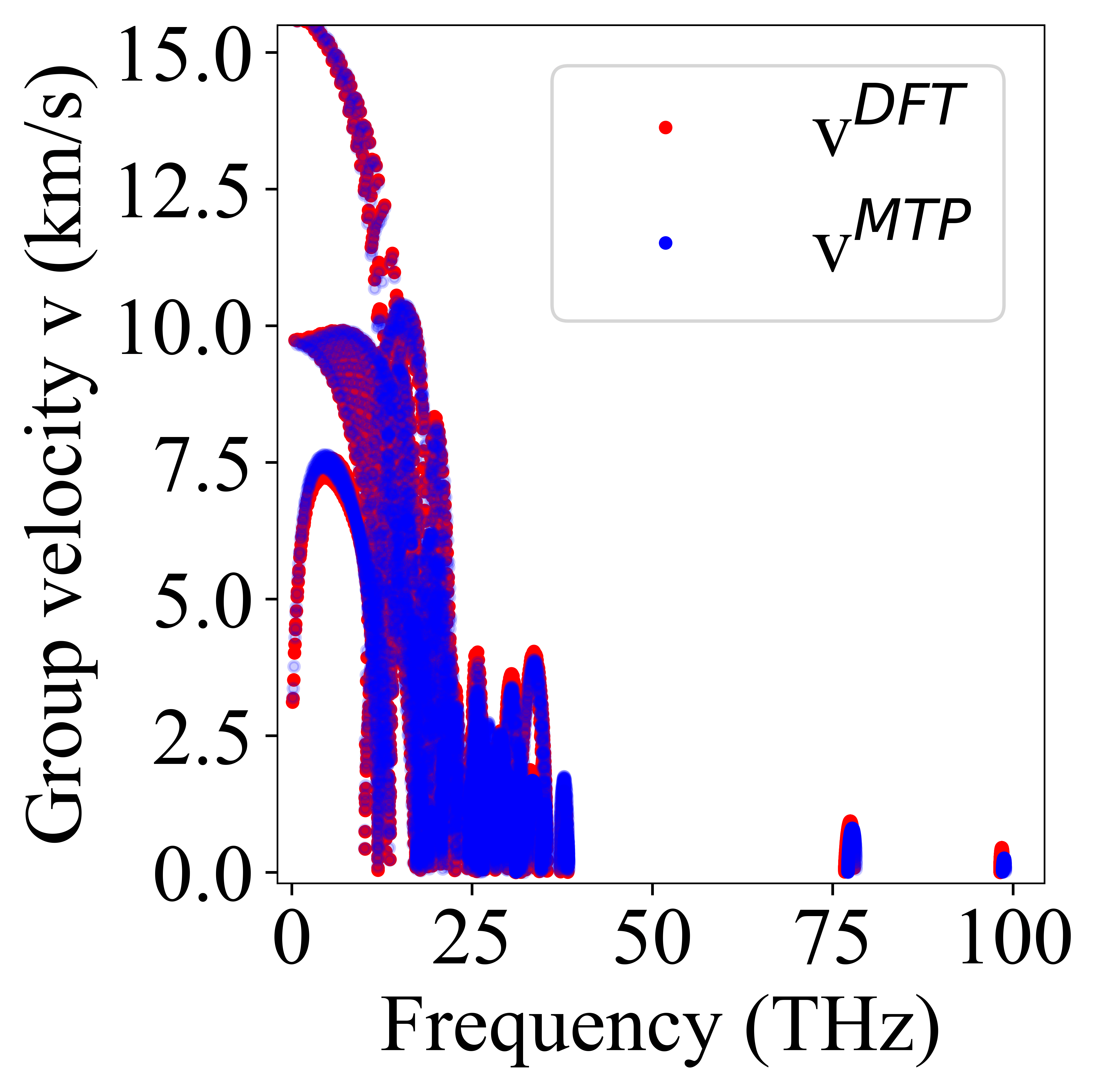}}
        (c)
	\end{minipage}
	\caption{Validation of MTP for BNnAB. Comparison of: (a) - forces obtained with DFT and MTP for 1000 snapshots from the trajectory of AIMD performed at 500 K (black dashed line represents an $x=y$ ideal linear fit), (b) - heat capacities and (c) - group velocities obtained with DFT and MTP in the harmonic approximation. }
	\label{fig:mtp_valid}
\end{figure*}

\begin{figure*}[h!]
	\begin{minipage}[h]{0.32\linewidth}
		\centering{\includegraphics[trim={0cm 0cm 0cm 0cm},clip, width=1\linewidth]{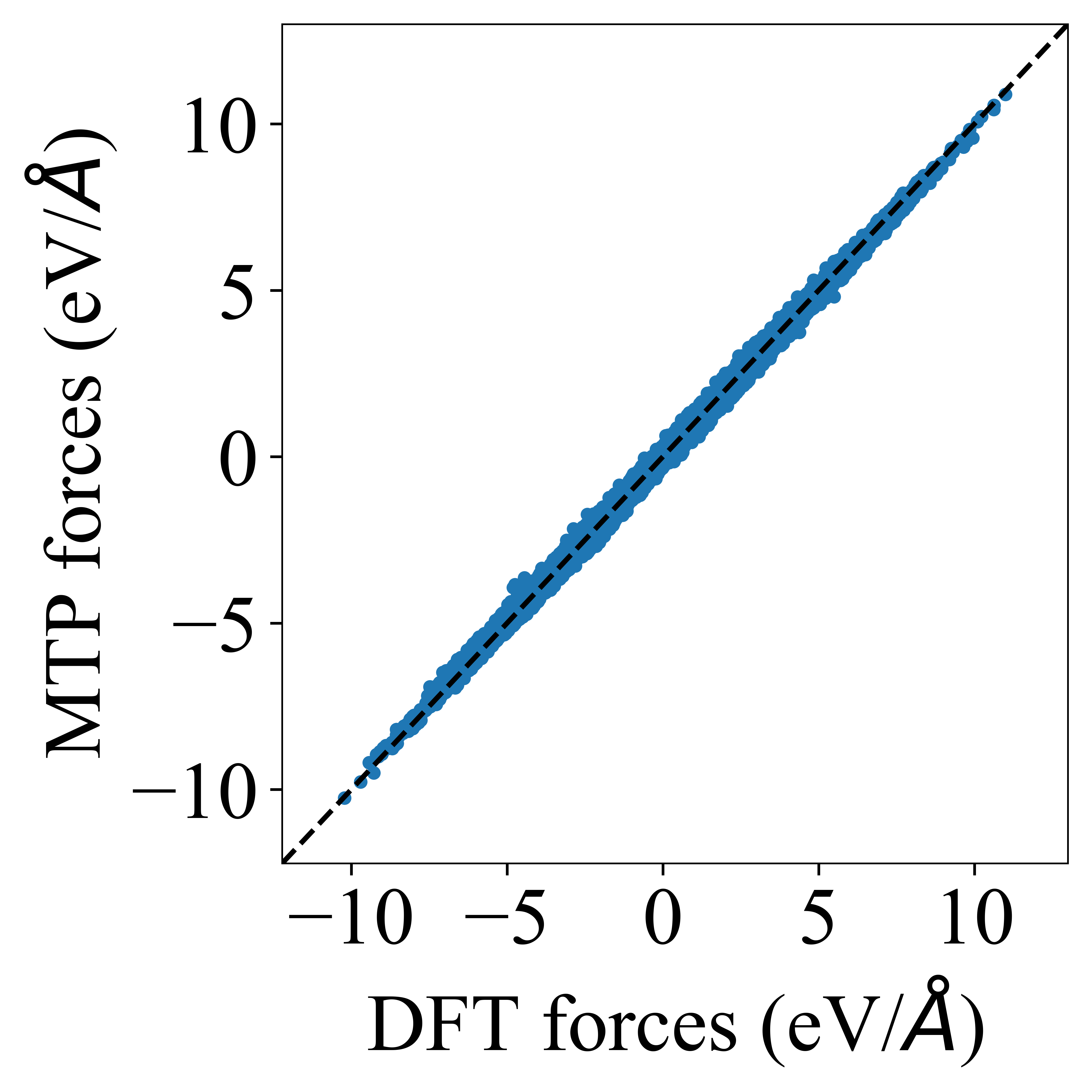}}
        (a)
	\end{minipage}%
 	\begin{minipage}[h]{0.32\linewidth}
		\centering{\includegraphics[trim={0cm 0cm 0cm 0cm},clip, width=1\linewidth]{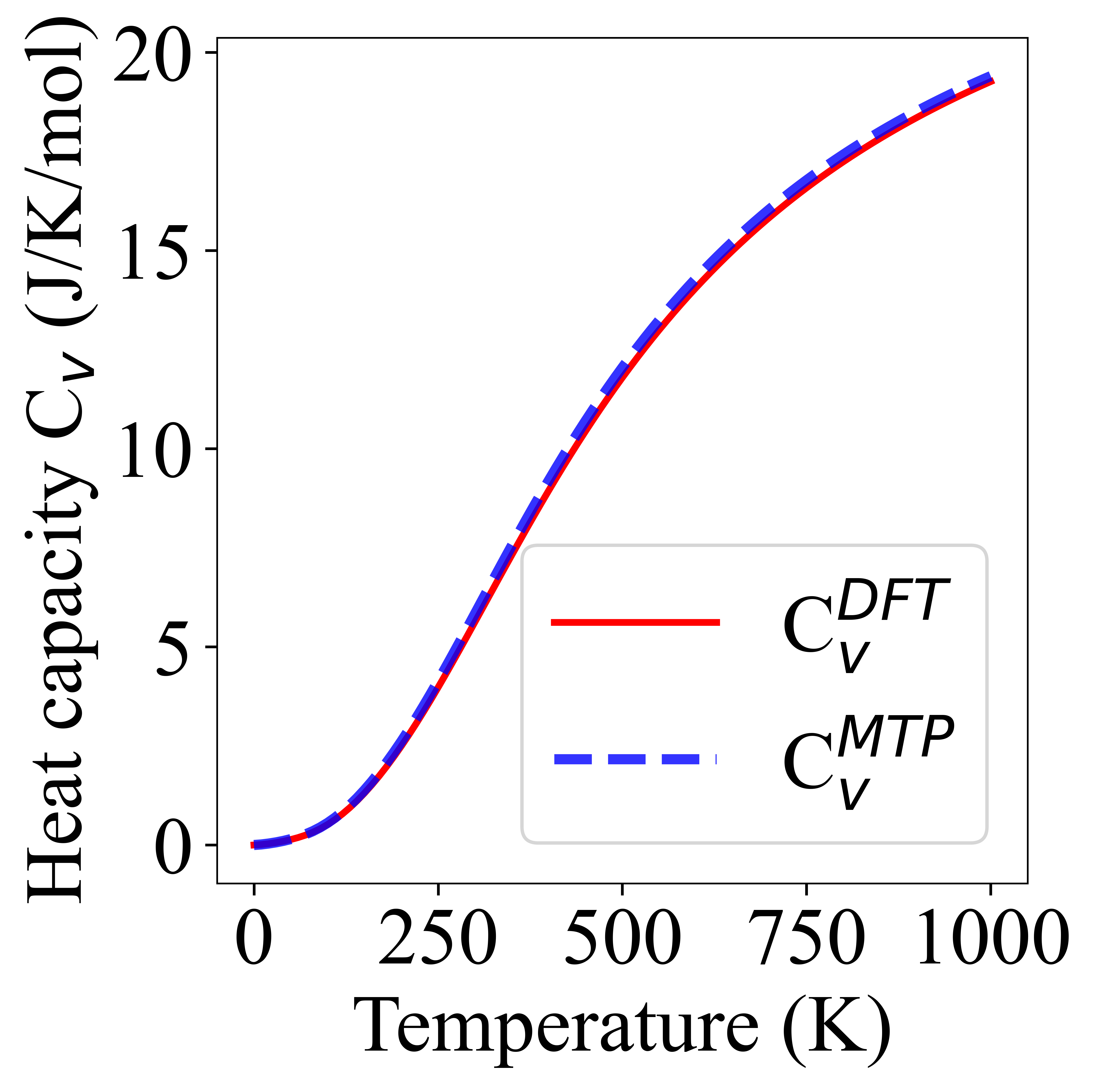}}
        (b)
	\end{minipage}%
    \begin{minipage}[h]{0.32\linewidth}
            \centering{\includegraphics[trim={0cm 0cm 0cm 0cm},clip, width=1\linewidth]{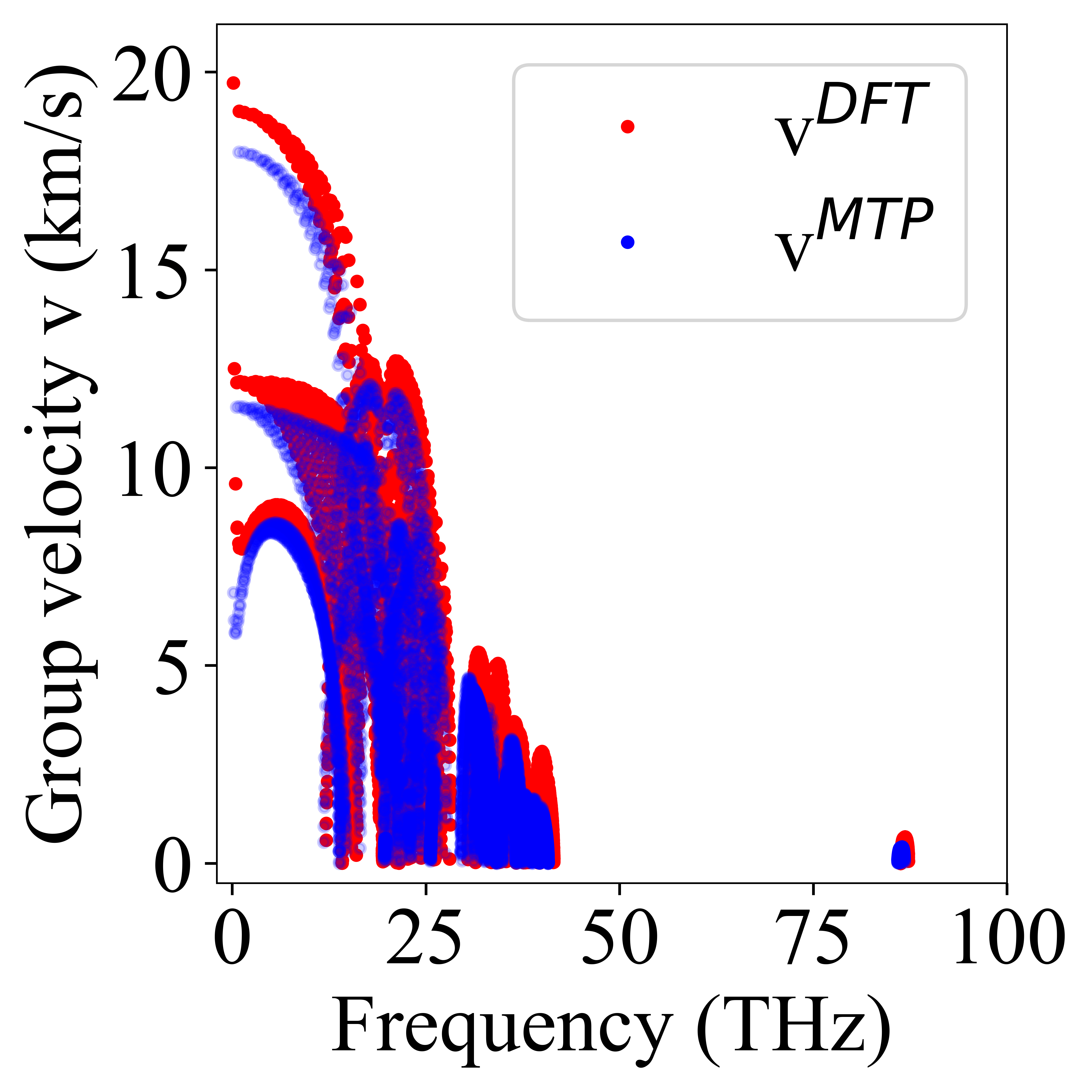}}
        (c)
	\end{minipage}
	\caption{Validation of MTP for DnAB. Comparison of: (a) - forces obtained with DFT and MTP for 1000 snapshots from the trajectory of AIMD performed at 1000 K (black dashed line represents an $x=y$ ideal linear fit), (b) - heat capacities and (c) - group velocities obtained with DFT and MTP in the harmonic approximation. }
	\label{fig:CH-mtp_valid}
\end{figure*}

\begin{figure*}[h!]

  	\begin{minipage}[h]{0.32\linewidth}
		\centering{\includegraphics[trim={0cm 0cm 0cm 0cm},clip, width=1\linewidth]{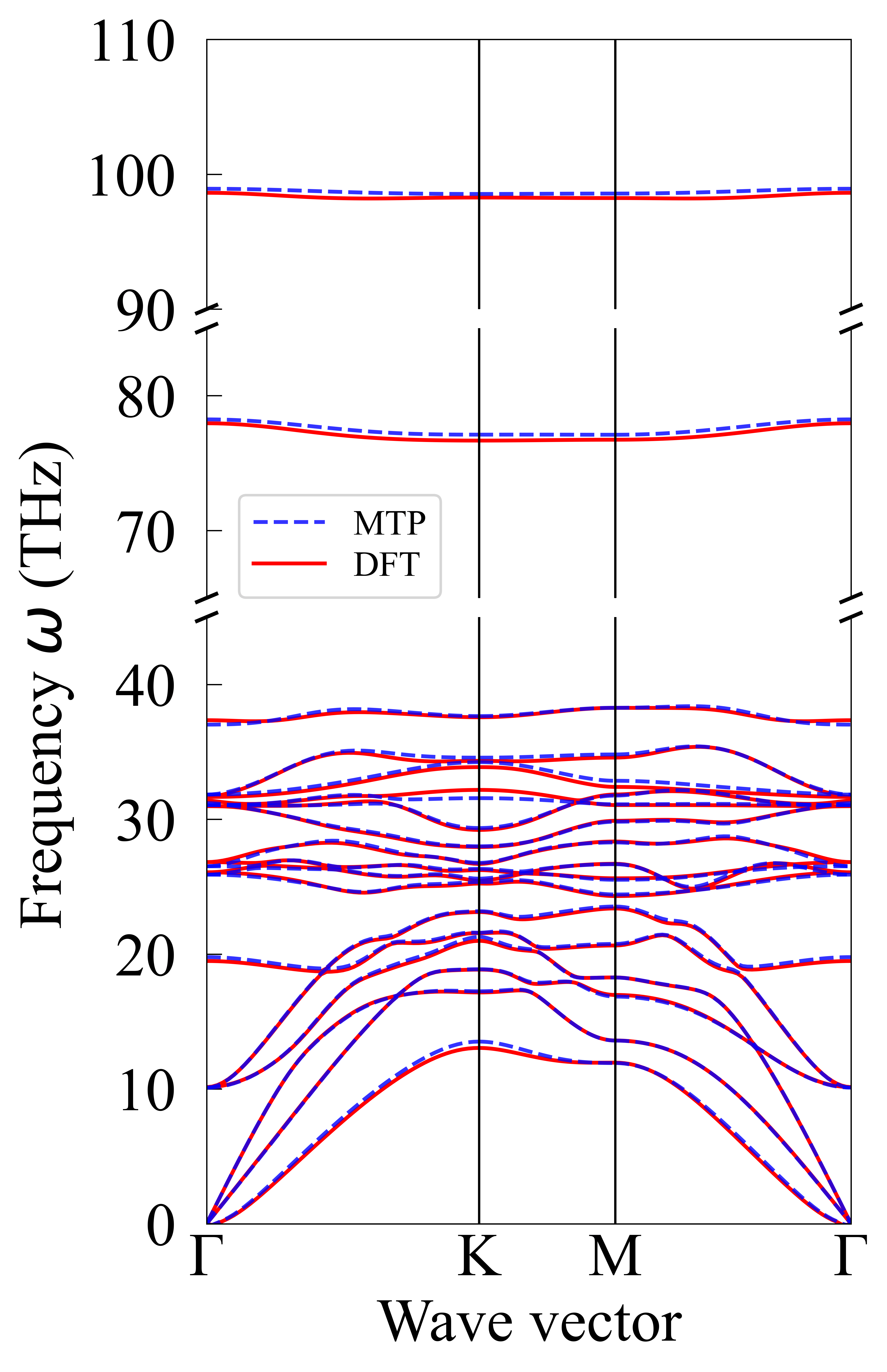}}
		(a)
          
	\end{minipage}
 	\begin{minipage}[h]{0.32\linewidth}
		\centering{\includegraphics[trim={0cm 0cm 0cm 0cm},clip, width=1\linewidth]{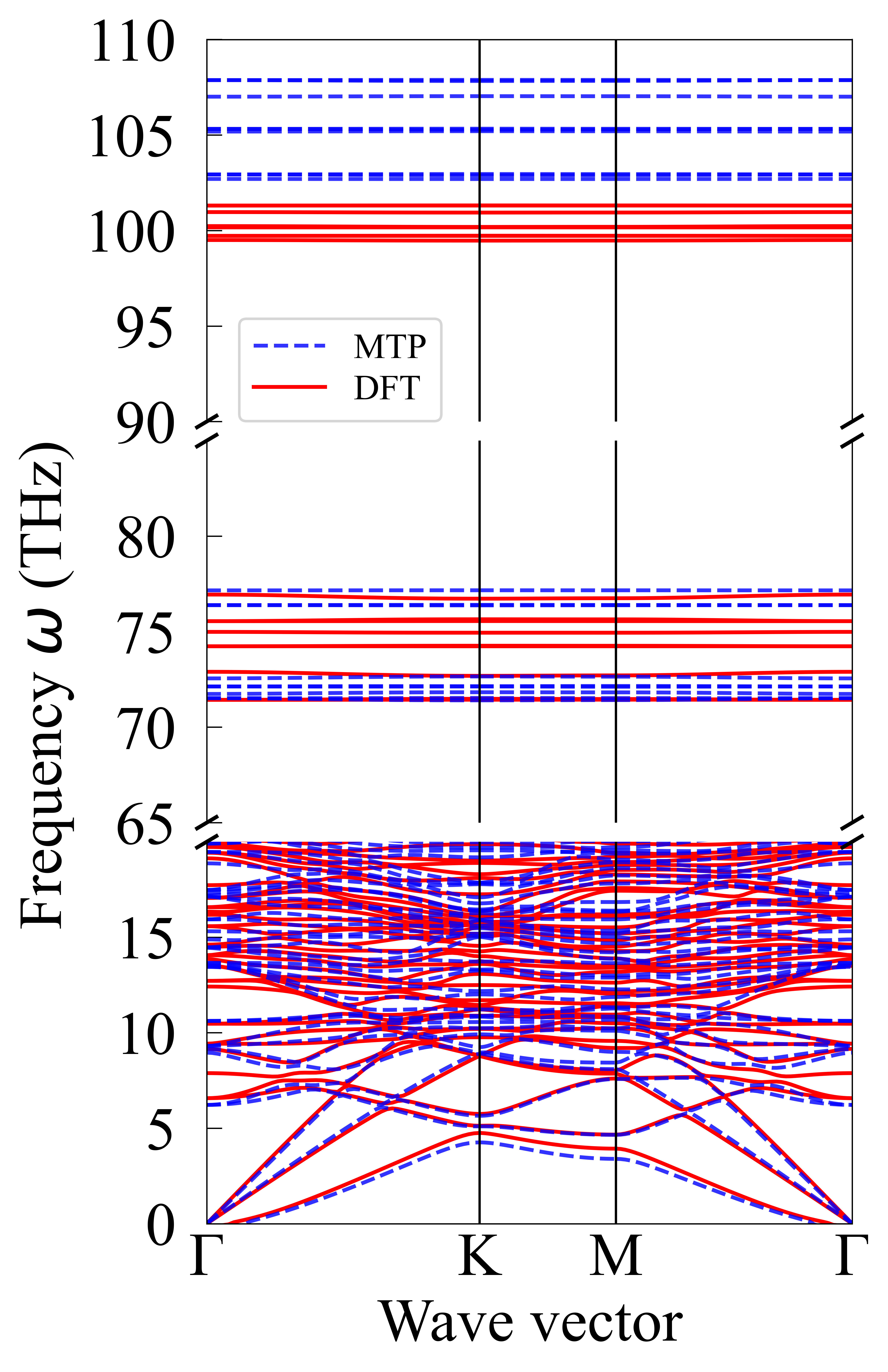}}
		(b)
         
	\end{minipage}
 	\begin{minipage}[h]{0.32\linewidth}
		\centering{\includegraphics[trim={0cm 0cm 0cm 0cm},clip, width=1\linewidth]{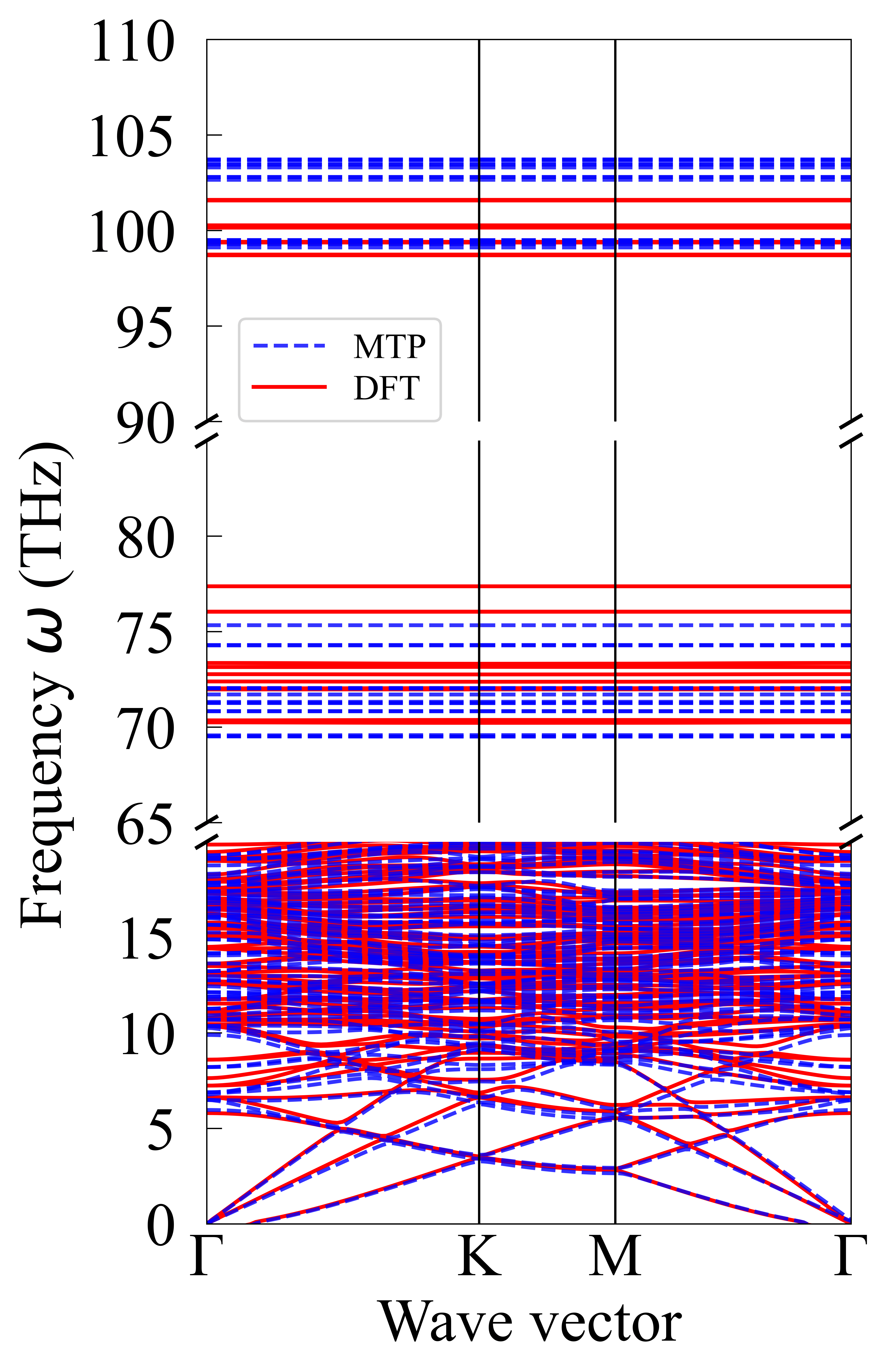}}
		(c)
          
	\end{minipage}
 
	\caption{Phonon band structures for the BNn$\theta$ Moir\'e lattices  ((a) - BNnAB, (b) - BNn21.8, (c) - BNn27.8) calculated with MTP and DFT. For convenience of visual perception, the frequency ranges are shown partially. The full graphs with uncut y axes can be found in Fig. S5.}
 
	\label{fig:phonons_21_27}
\end{figure*}

\begin{figure*}[h!]

  	\begin{minipage}[h]{0.32\linewidth}
		\centering{\includegraphics[trim={0cm 0cm 0cm 0cm},clip, width=1\linewidth]{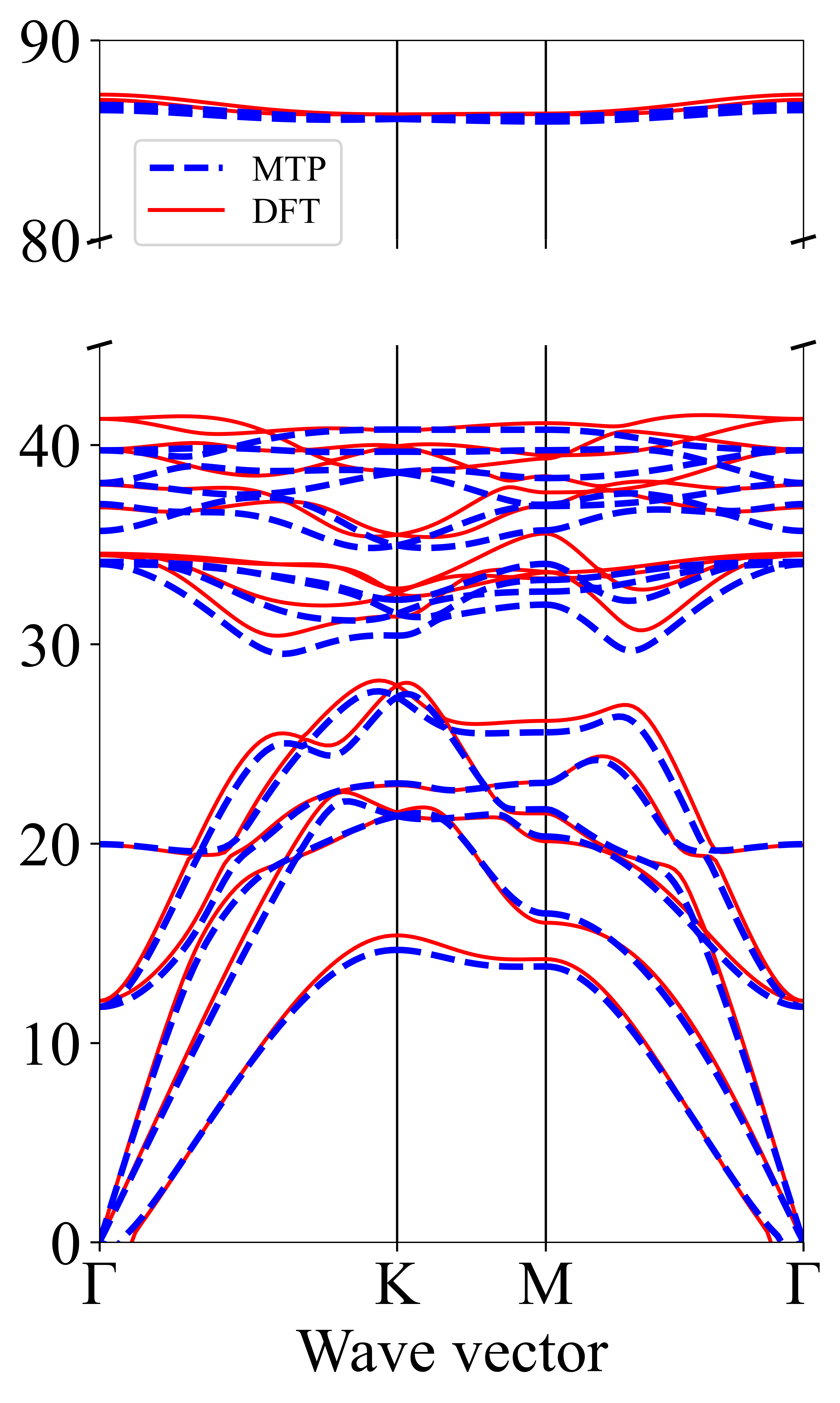}}
          
	\end{minipage}
 	\begin{minipage}[h]{0.32\linewidth}
		\centering{\includegraphics[trim={0cm 0cm 0cm 0cm},clip, width=1\linewidth]{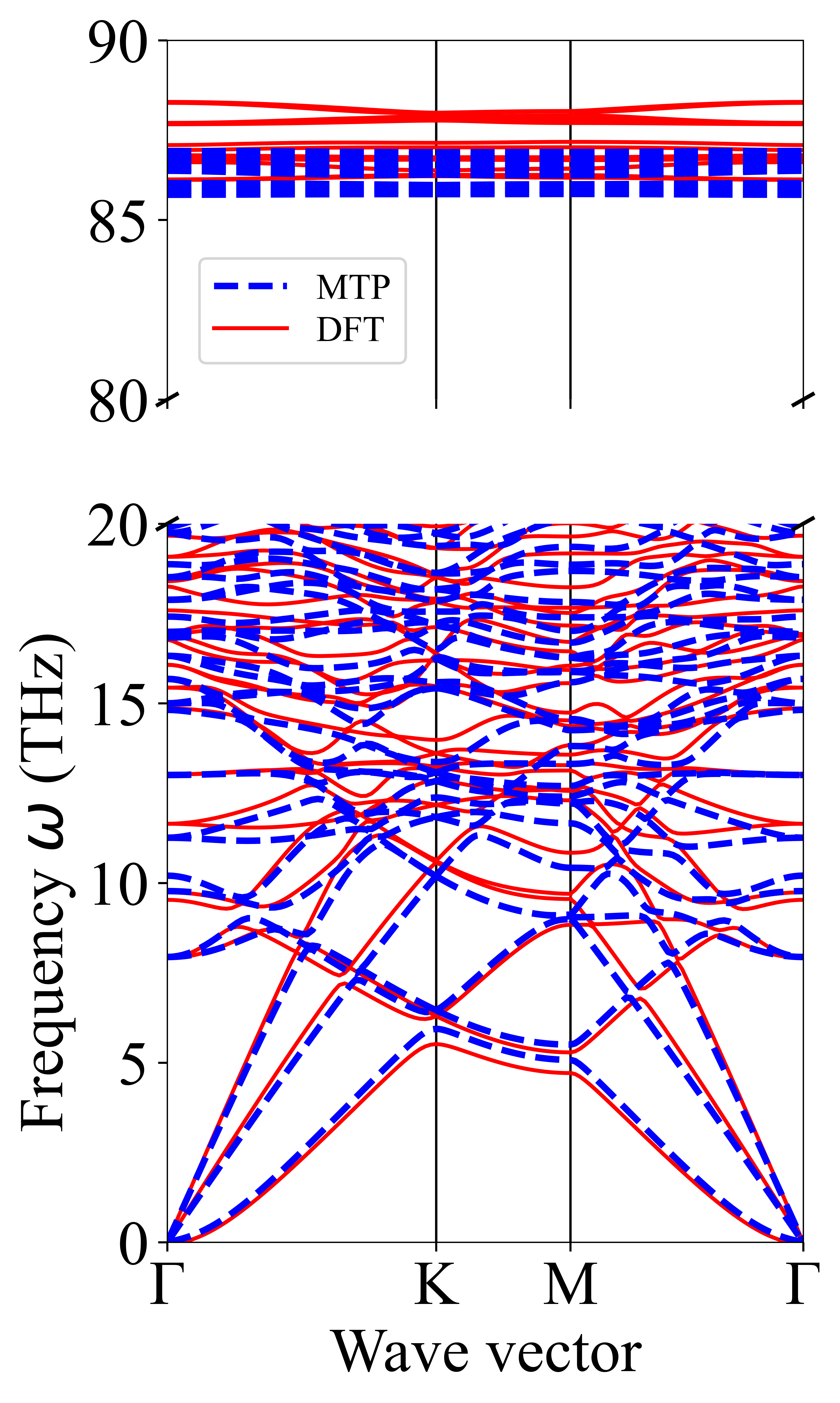}}
         
	\end{minipage}
 	\begin{minipage}[h]{0.32\linewidth}
		\centering{\includegraphics[trim={0cm 0cm 0cm 0cm},clip, width=1\linewidth]{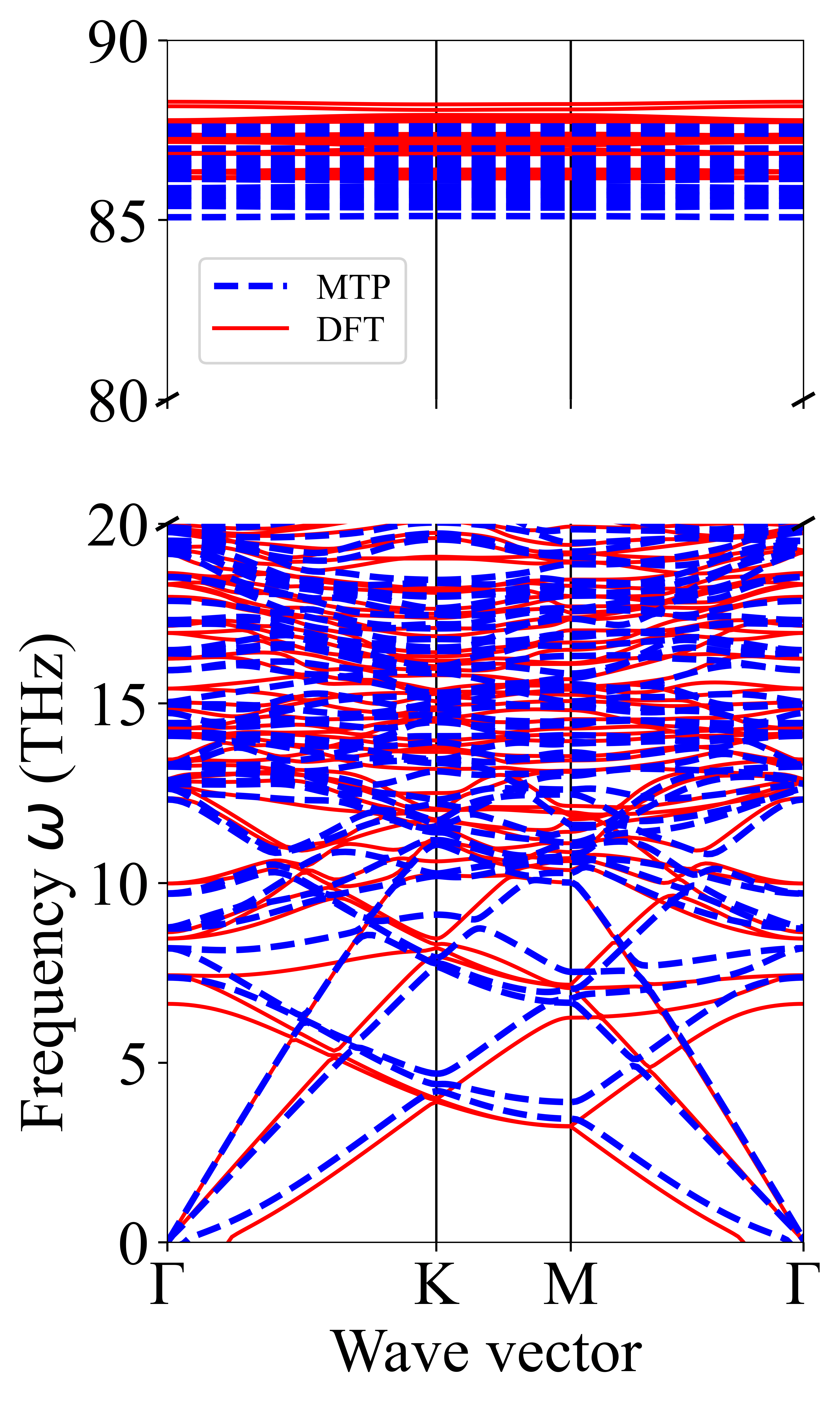}}
          
	\end{minipage}
 
	\caption{Phonon band structures for the Dn$\theta$  Moir\'e lattices  ((a) - DnAB, (b) - Dn21.8, (c) - Dn27.8) calculated with MTP and DFT. For convenience of visual perception, the frequency ranges are shown partially. The full graphs with uncut y axes can be found in Fig. S6.}
 
	\label{fig:CH-phonons_21_27}
\end{figure*}

The MTP was validated by comparing the forces predicted by the MTP for the AIMD trajectory snapshots with the values obtained by the DFT-AIMD simulation. The phonon band structures, group velocities, and heat capacities calculated with DFT and MTP for all Dn$\theta$ and BNn$\theta$ Moir\'e diamanes were also compared.  
The results are shown in Figs.~\ref{fig:mtp_valid} - \ref{fig:CH-phonons_21_27}  and Figs. S14 - S17, where one can clearly see that the MTP reproduces the forces and phonon properties remarkably well, demonstrating excellent agreement with the DFT results. Thus, the actively trained MTPs correctly describe the interatomic interactions in the BNn$\theta$ and Dn$\theta$ structures.

\subsection{Lattice thermal conductivity calculations}

The thermal conductivity of any material comprises electronic and lattice (phonon) contributions. In semiconducting materials such as the BNn$\theta$ and Dn$\theta$ Moir\'e lattices studied here, the electronic contribution is negligible, making the lattice component dominant. To obtain the lattice thermal conductivity in the BTE-based approach according to Eq. (\ref{eq:kappa}), the calculations of phonon heat capacities, group velocities, and lifetimes are required. All these values are obtained from the phonon band structures calculated with the MTP that are shown in Figs.~\ref{fig:phonons_21_27} and \ref{fig:CH-phonons_21_27}.

We investigate and compare the phonon properties in the 0-45 THz range because this region contributes mostly to the LTC. As shown in Figs. \ref{fig:mtp_valid}(d) and \ref{fig:CH-mtp_valid}(d), phonon group velocities are significantly higher in this range, and thus, the phonons from this region make the main contribution to the LTC (Eq. (\ref{eq:kappa})). Therefore, analyzing phonon properties (heat capacities, group velocities, and lifetimes) within 0-45 THz is sufficient to explain the dependence of LTC on the twist angle.

\begin{figure*}[h!]
\centering
	\begin{minipage}[h]{0.45\linewidth}
		\centering{\includegraphics[trim={0cm 0cm 0cm 0cm},clip, width=1\linewidth]{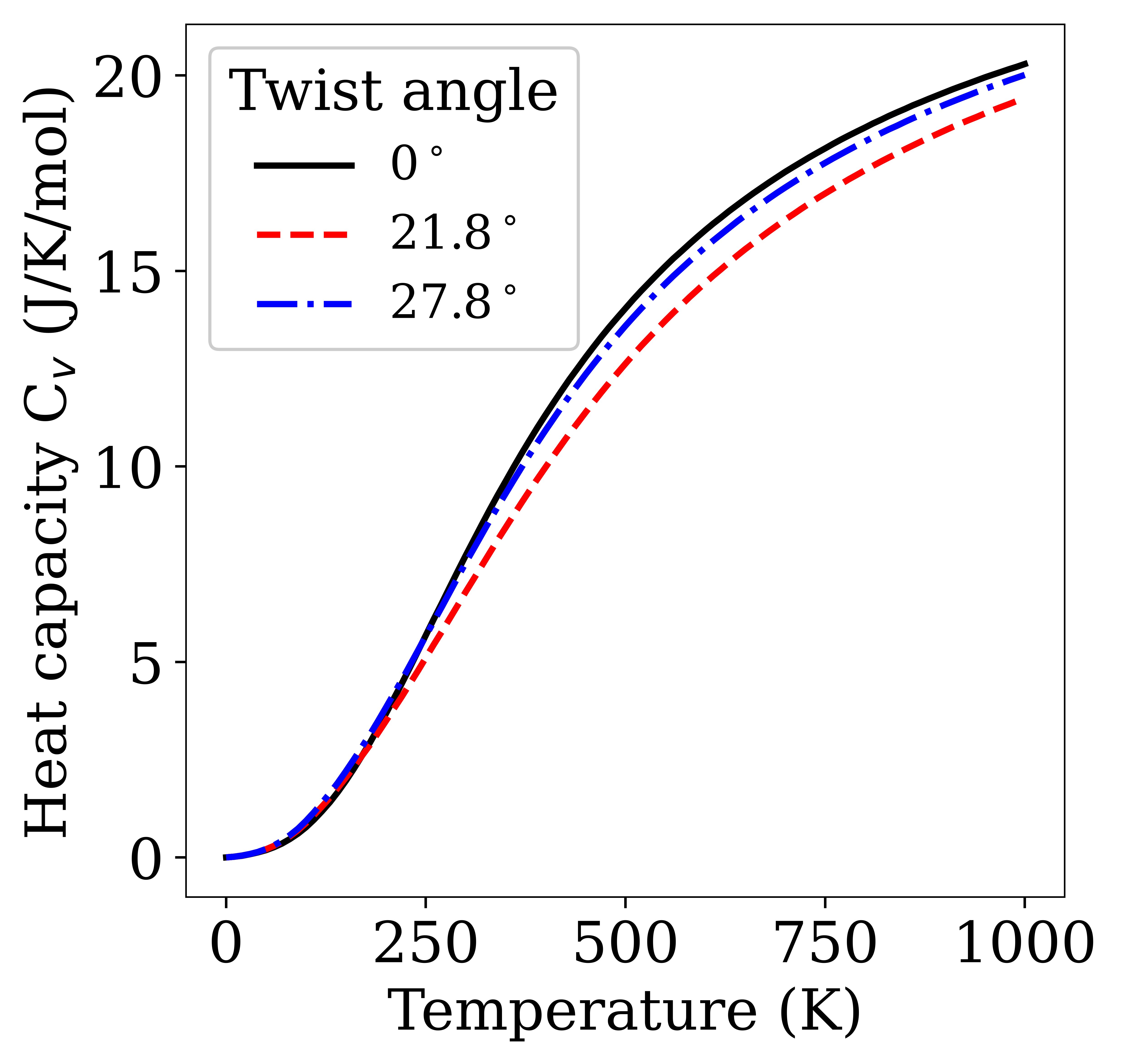}}
(a)
	\end{minipage}%
        \begin{minipage}[h]{0.45\linewidth}
		\centering{\includegraphics[trim={0cm 0cm 0cm 0cm},clip, width=1\linewidth]{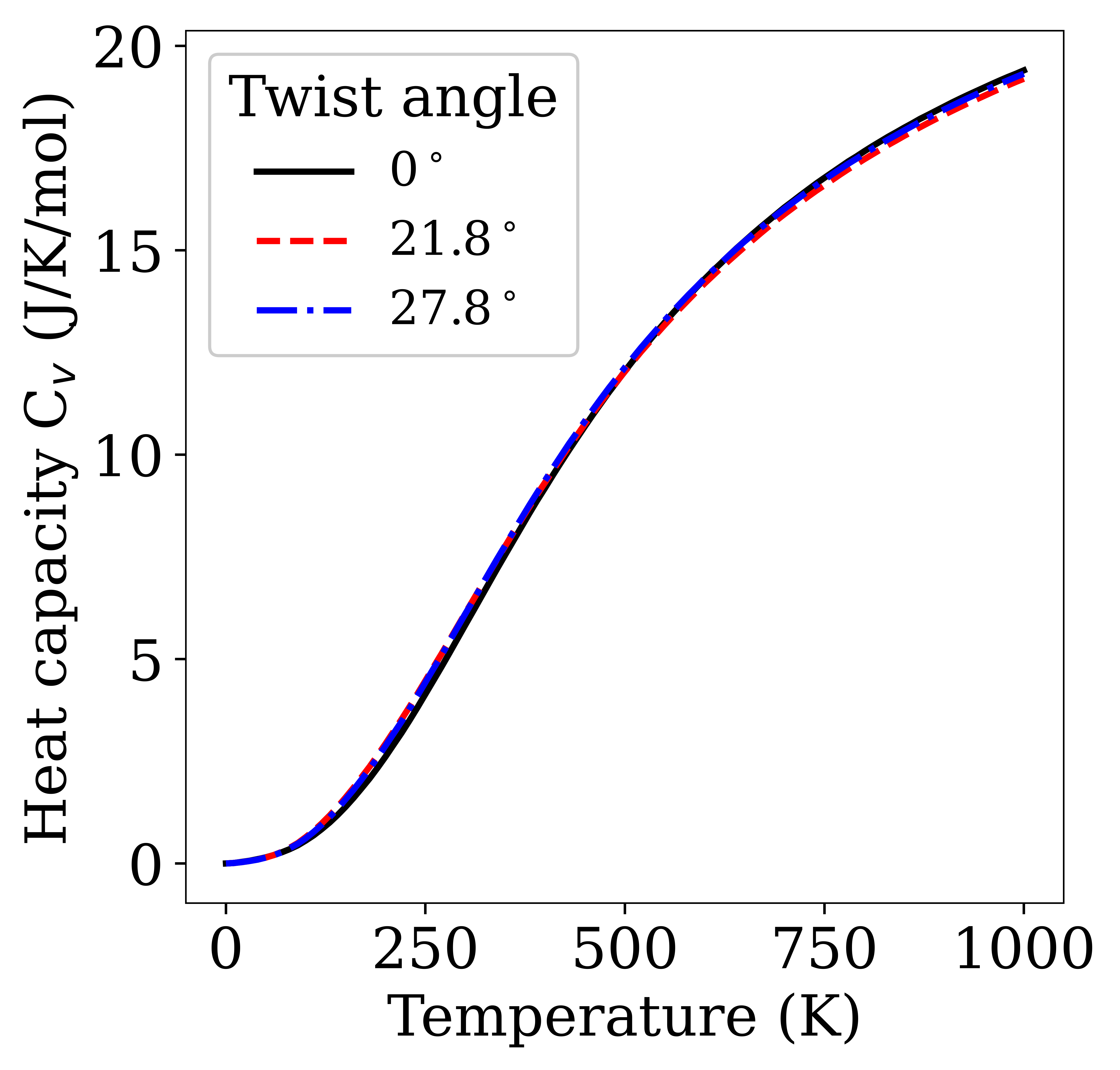}}
(b)
	\end{minipage}
	\caption{Phonon heat capacities in BNn$\theta$ (a) and Dn$\theta$ (b) structures. }
	\label{fig:c_q_v_bnh}
\end{figure*}

\begin{figure*}[h!]
    \begin{minipage}[h]{0.5\linewidth}
		\centering{\includegraphics[trim={0cm 0cm 0cm 0cm},clip, width=1\linewidth]{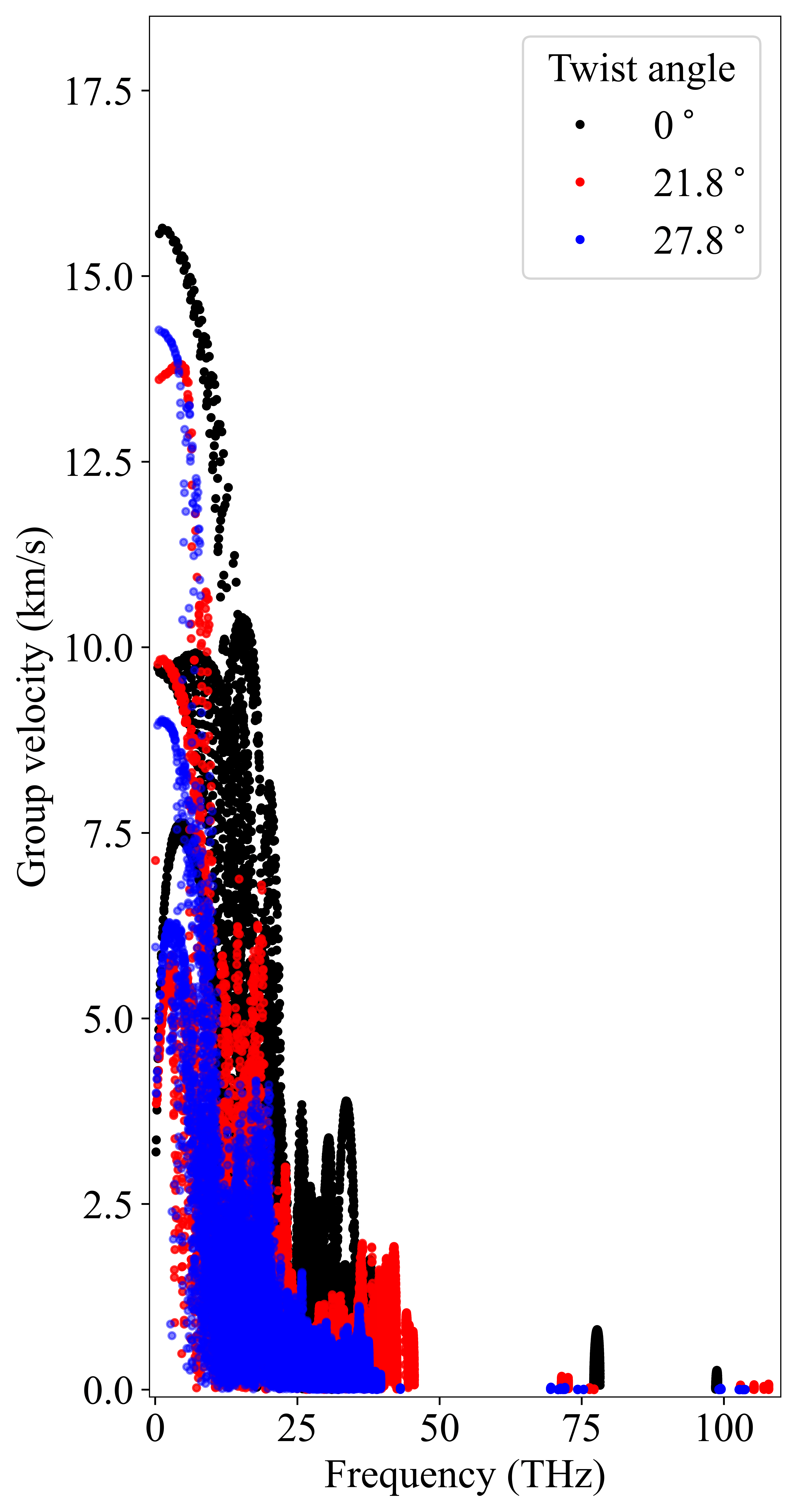}}
        (a)

	\end{minipage}%
	\begin{minipage}[h]{0.5\linewidth}
		\centering{\includegraphics[trim={0cm 0cm 0cm 0cm},clip, width=1\linewidth]{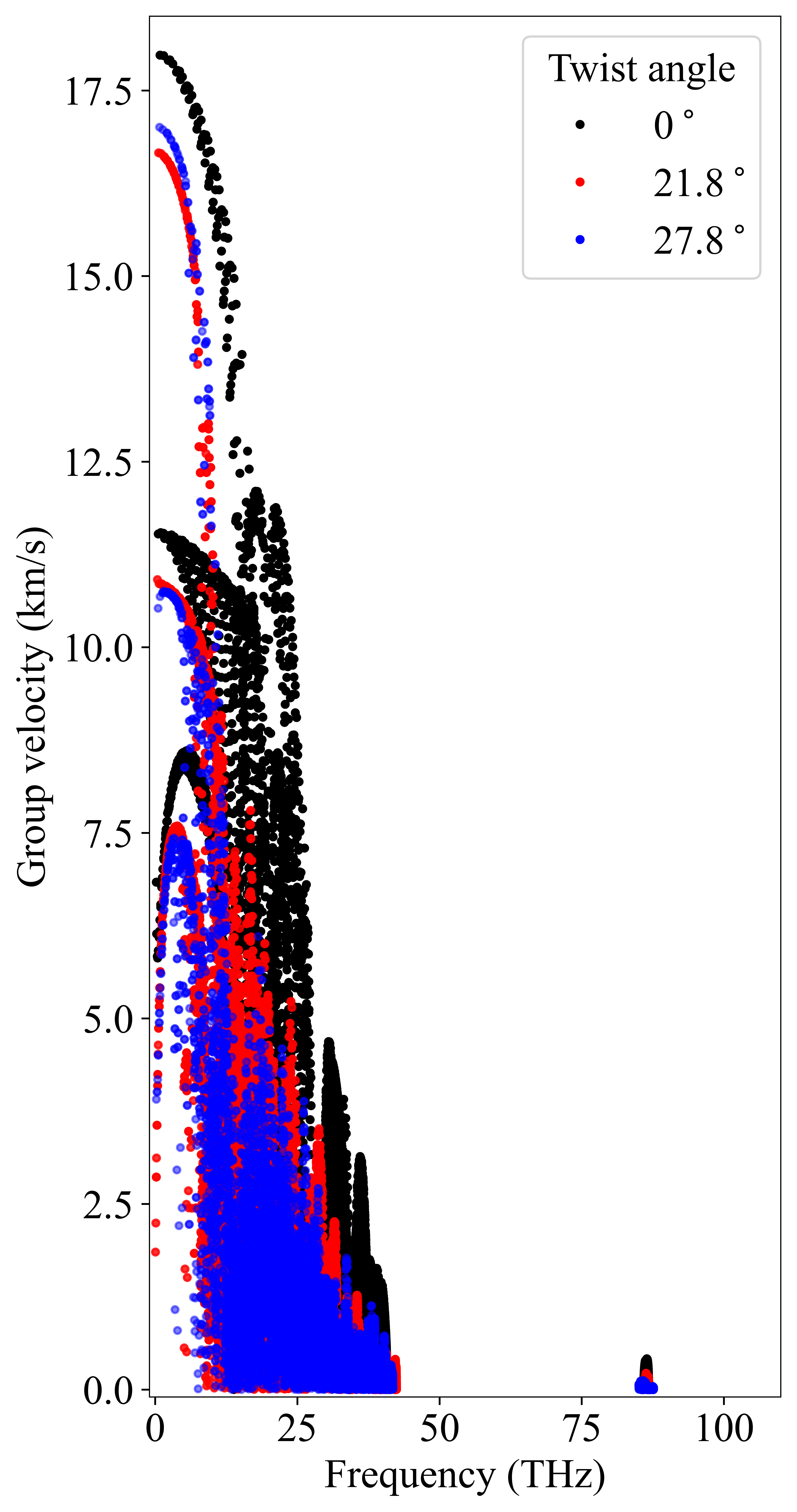}}
        (b)

	\end{minipage}

	\caption{Phonon group velocities in BNn$\theta$ (a) and Dn$\theta$ (b) structures.}
	\label{fig:Gv_bnh}
\end{figure*}

As shown in Fig.~\ref{fig:c_q_v_bnh}, the values of the heat capacities are similar for all three Moir\'e lattices in both BNn$\theta$ and Dn$\theta$ cases. According to the data given in Figs.~\ref{fig:Gv_bnh} and S2, the phonon group velocities also have the same order of magnitude for the three Moir\'e lattices of hydrogenated BN and graphene bilayers.

\begin{figure*}[h]
	\centering
    	\begin{minipage}[h]{0.45\linewidth}
		\centering{\includegraphics[trim={0cm 0cm 0cm 0cm},clip, width=1\linewidth]{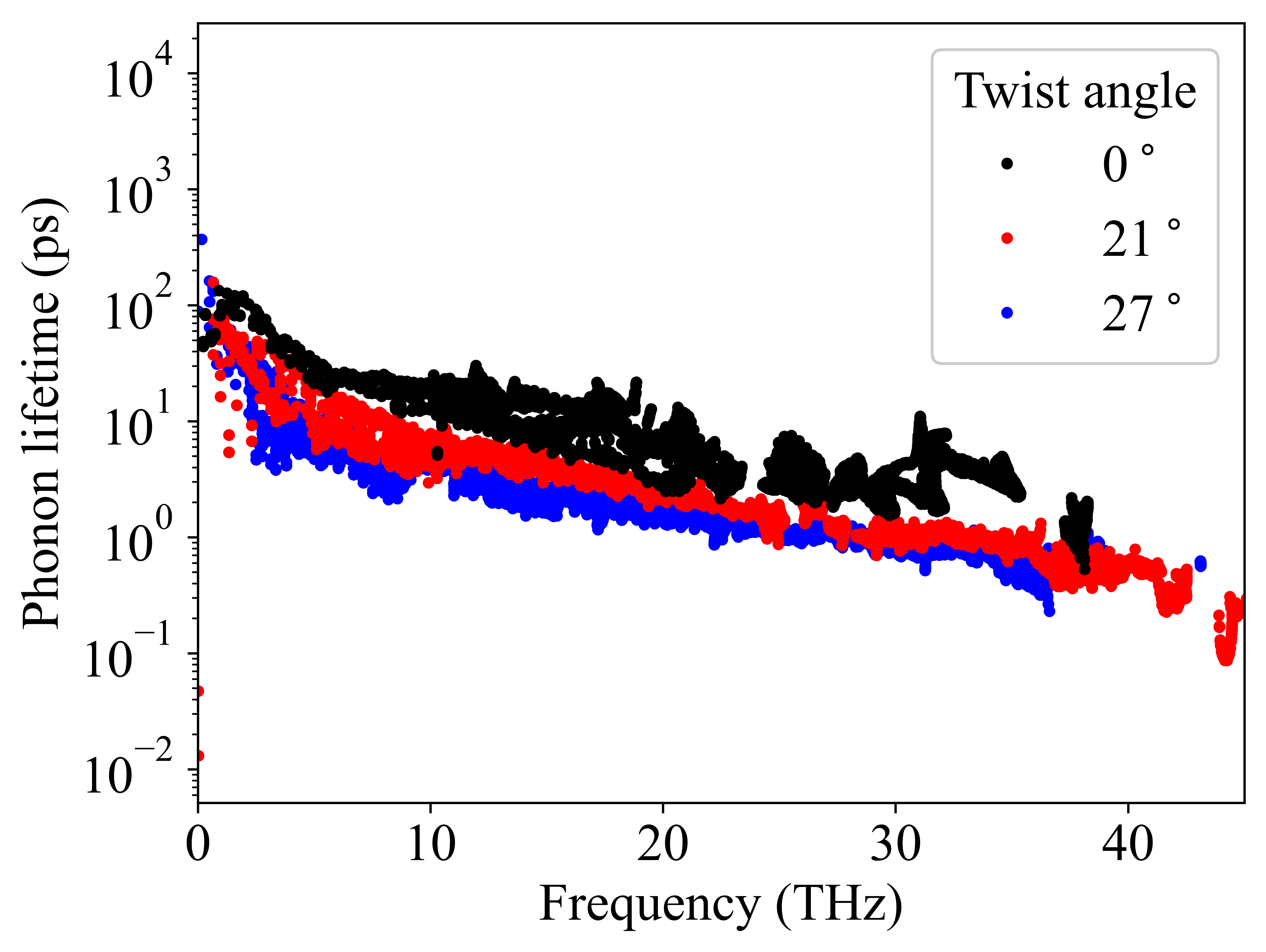}}
(a)
	\end{minipage}%
        \begin{minipage}[h]{0.45\linewidth}
		\centering{\includegraphics[trim={0cm 0cm 0cm 0cm},clip, width=1\linewidth]{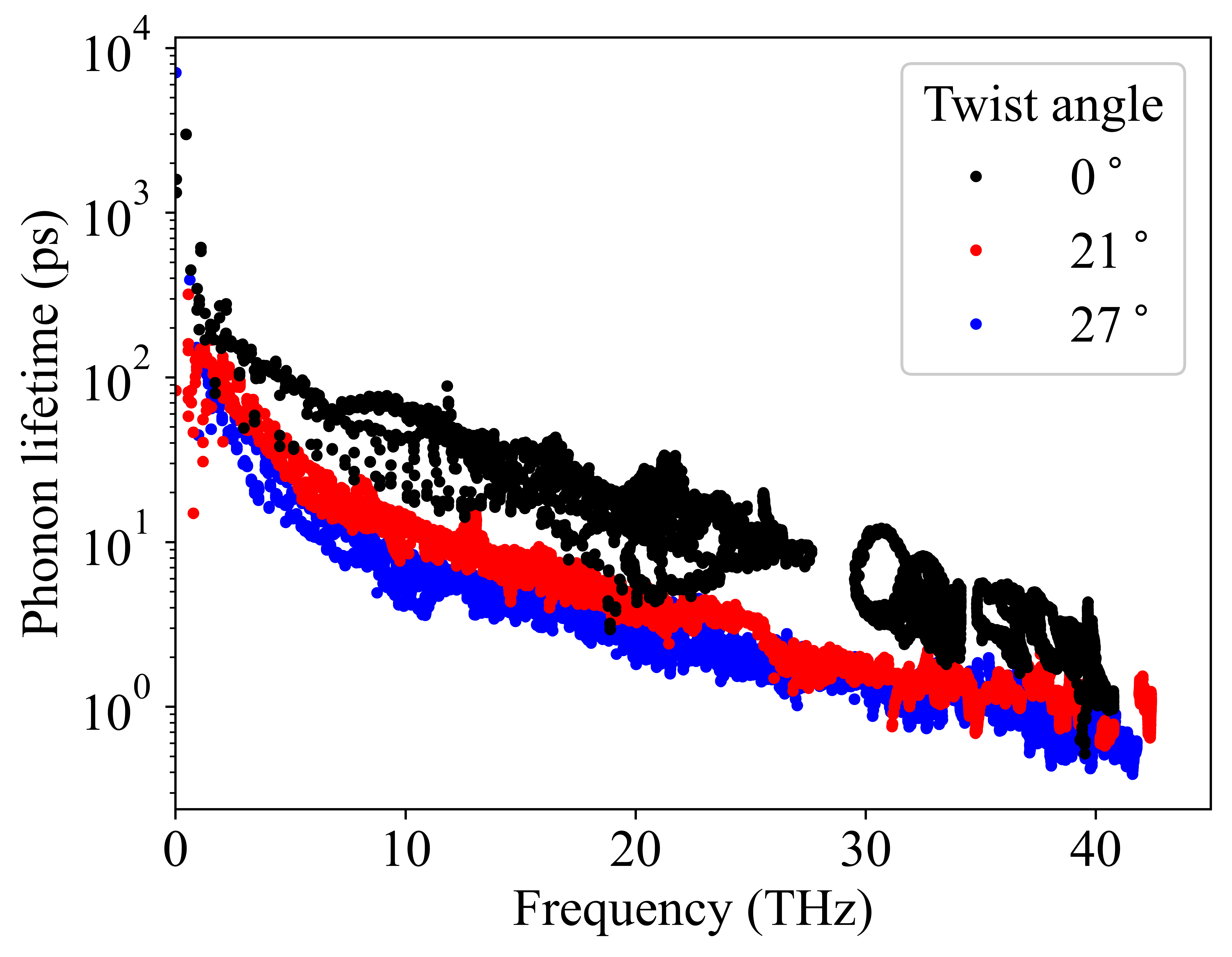}}
(b)
	\end{minipage}

	\caption{Phonon lifetimes for BNn$\theta$ (a) and Dn$\theta$ (b) structures at T = 300 K.}
	\label{fig:lifetime}
\end{figure*}

Another important characteristic that was calculated is the dependence of phonon lifetimes on temperature and the twist angle $\theta$. It affects the efficiency of heat transport through the material, i.e., shorter lifetimes can lead to lower thermal conductivity. As shown in Fig.~\ref{fig:lifetime}, the phonon lifetimes for the BNnAB and DnAB are higher by up to one order of magnitude than those for the two other structures at T = 300 K. The similar difference between phonon lifetimes in twisted and untwisted structures of BNn$\theta$ and Dn$\theta$ is observed at higher temperatures (400-700 K) (see Supplementary Information, Figs. S3 and S4).

\begin{figure*}[h]
	\begin{minipage}[h]{0.32\linewidth}
		\centering{\includegraphics[trim={0cm 0cm 0cm 0cm},clip, width=1\linewidth]{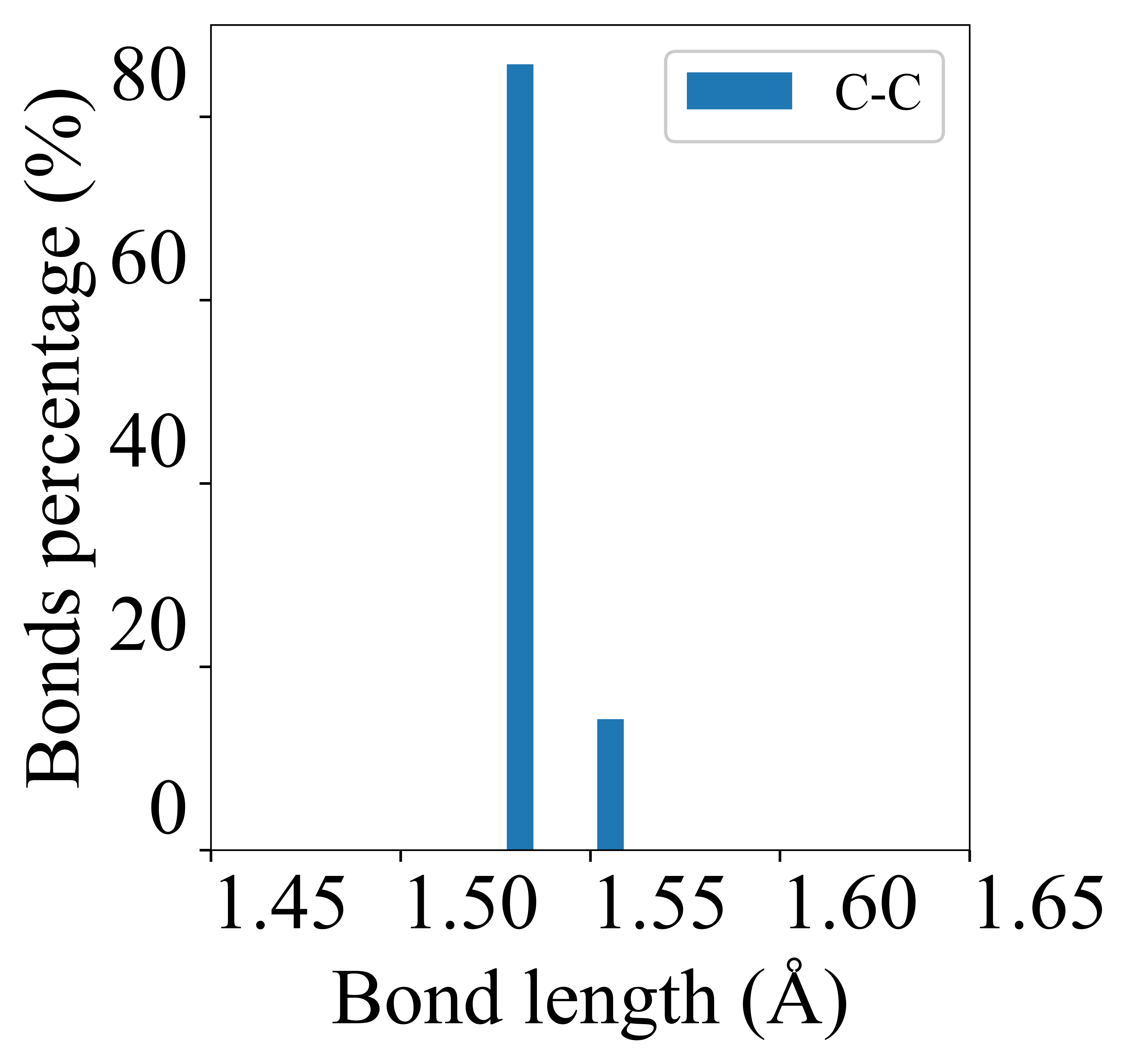}}
        (a)
	\end{minipage}%
 	\begin{minipage}[h]{0.32\linewidth}
		\centering{\includegraphics[trim={0cm 0cm 0cm 0cm},clip, width=1\linewidth]{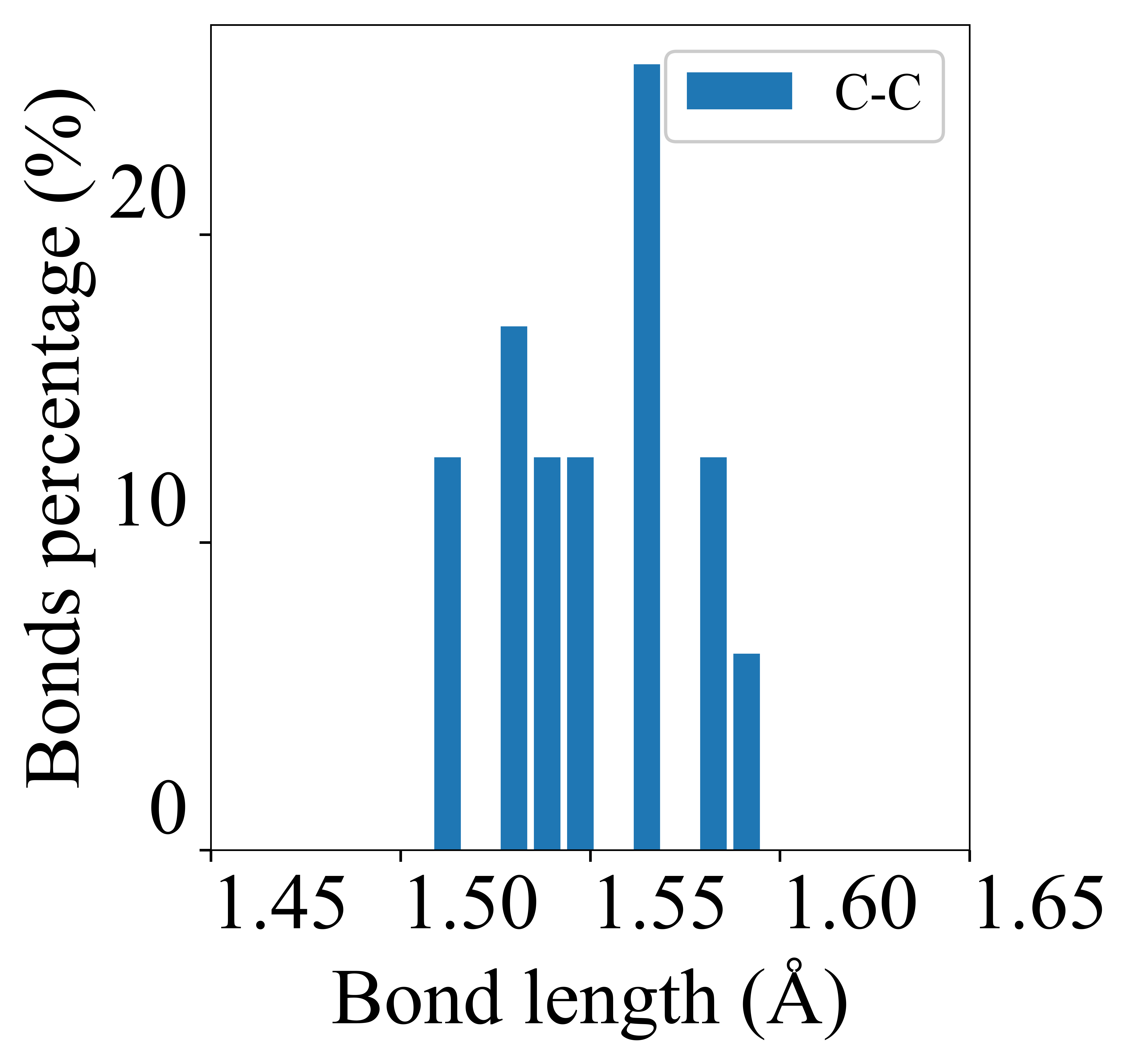}}
        (b)
	\end{minipage}%
    \begin{minipage}[h]{0.32\linewidth}
            \centering{\includegraphics[trim={0cm 0cm 0cm 0cm},clip, width=1\linewidth]{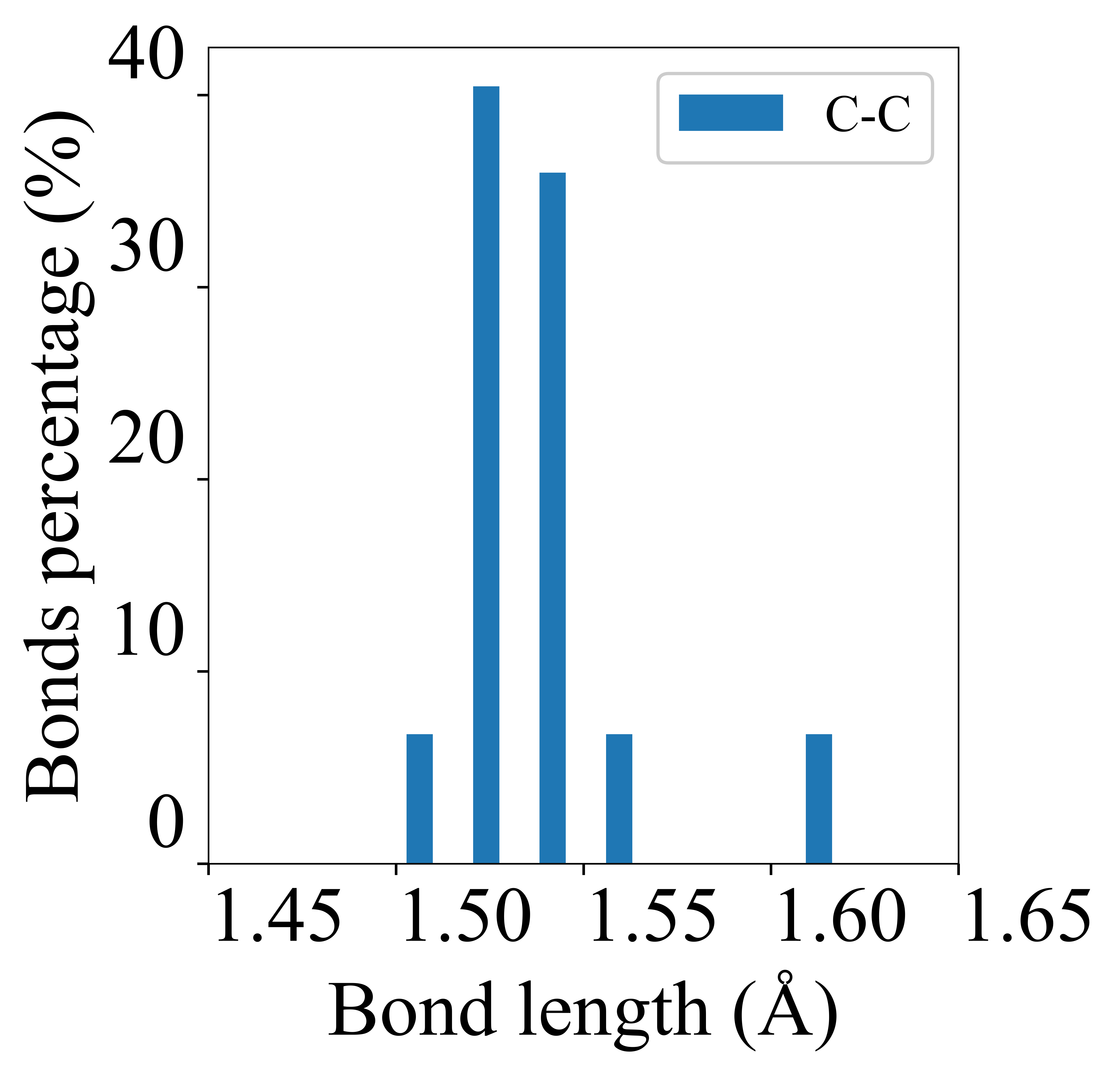}}
        (c)
	\end{minipage}

	\caption{C-C bond length distributions for the hydrogenated graphene Moir\'e lattices (twist angles: (a) - 0$^\circ$, (b) - 21.8$^\circ$, (c) - 27.8$^\circ$).}
	\label{fig:bond-len-ch}
\end{figure*}

\begin{figure*}[h]
	\begin{minipage}[h]{0.32\linewidth}
		\centering{\includegraphics[trim={0cm 0cm 0cm 0cm},clip, width=1\linewidth]{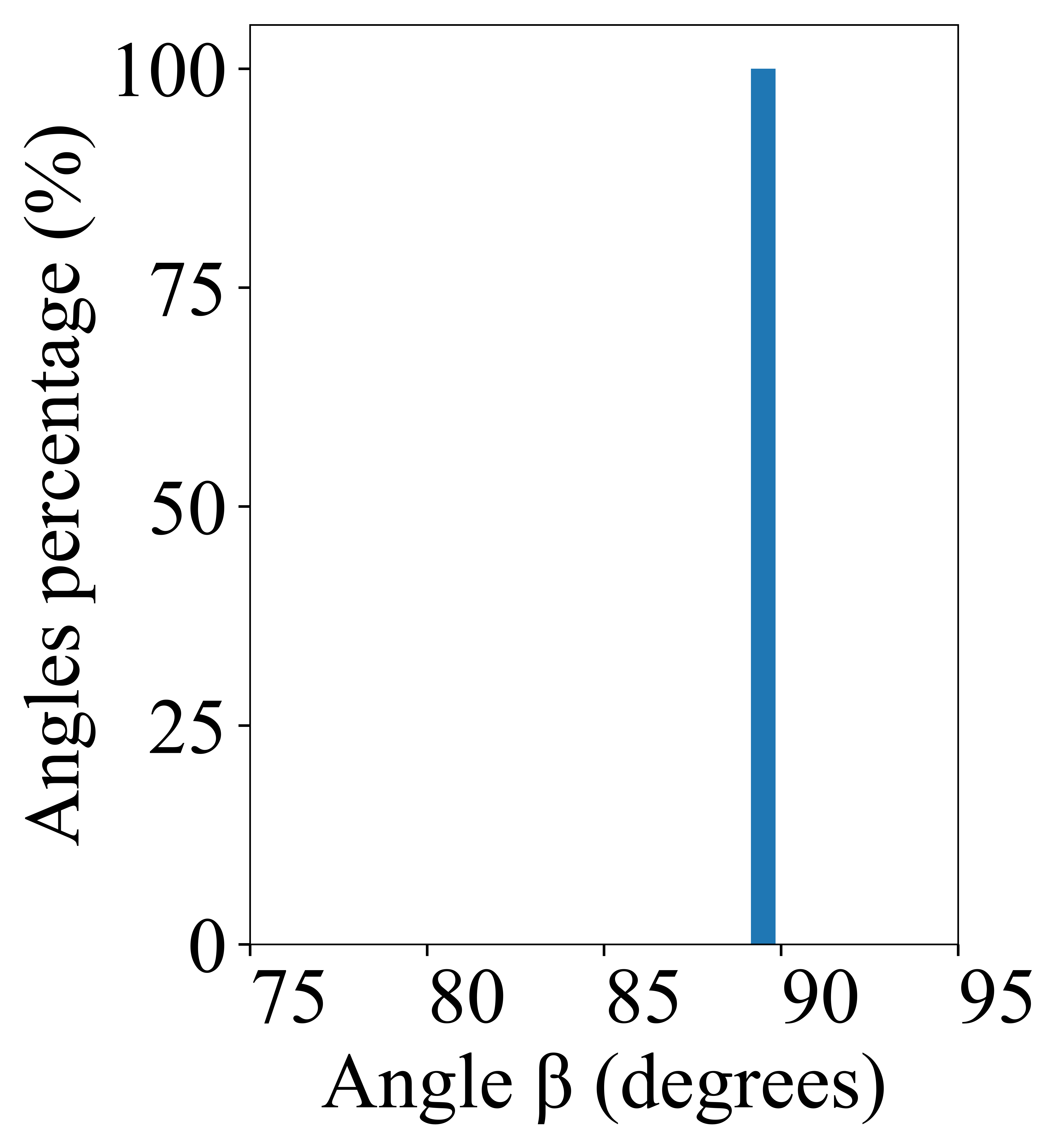}}
        (a)
	\end{minipage}%
 	\begin{minipage}[h]{0.32\linewidth}
		\centering{\includegraphics[trim={0cm 0cm 0cm 0cm},clip, width=1\linewidth]{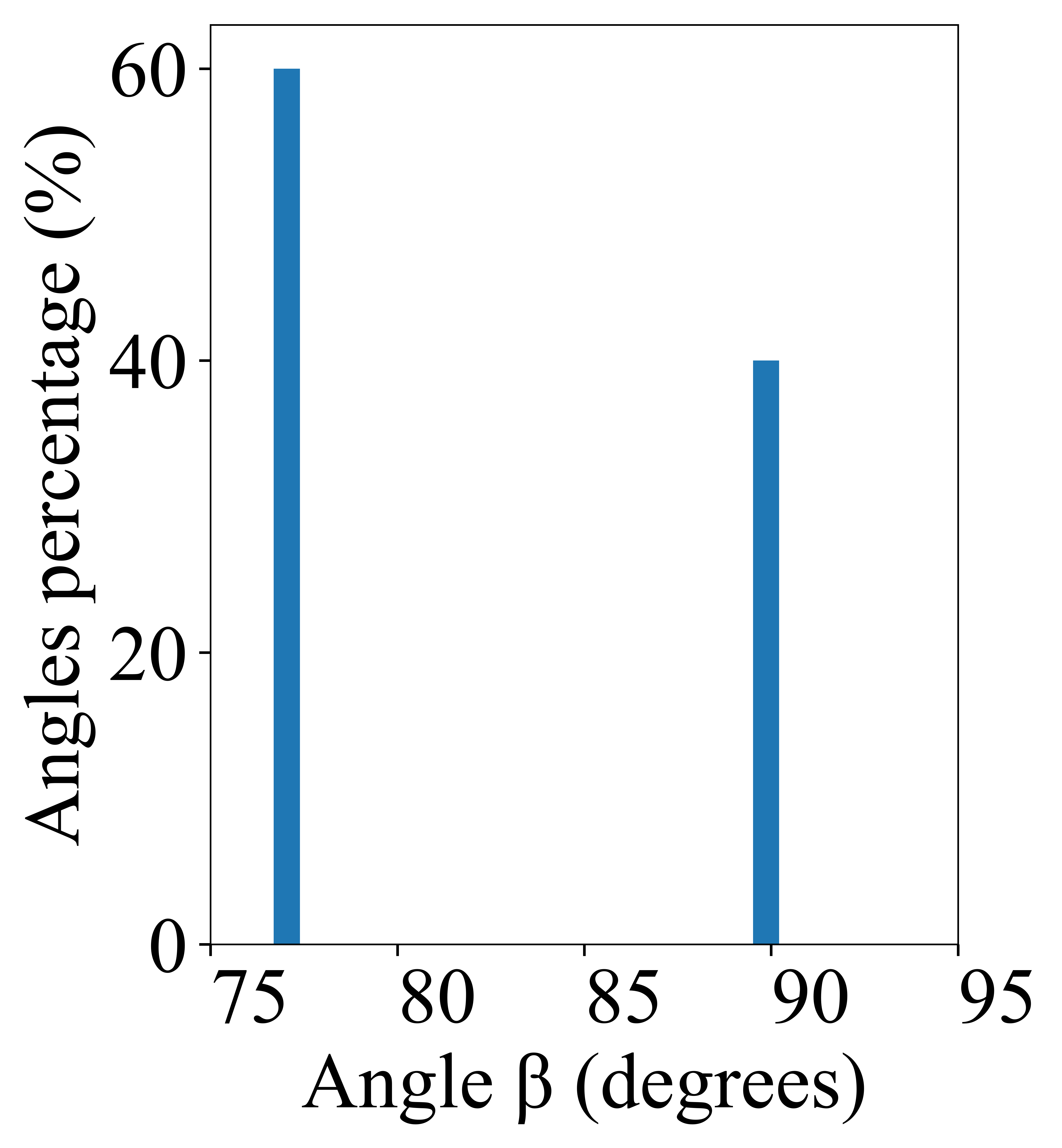}}
        (b)
	\end{minipage}%
    \begin{minipage}[h]{0.32\linewidth}
            \centering{\includegraphics[trim={0cm 0cm 0cm 0cm},clip, width=1\linewidth]{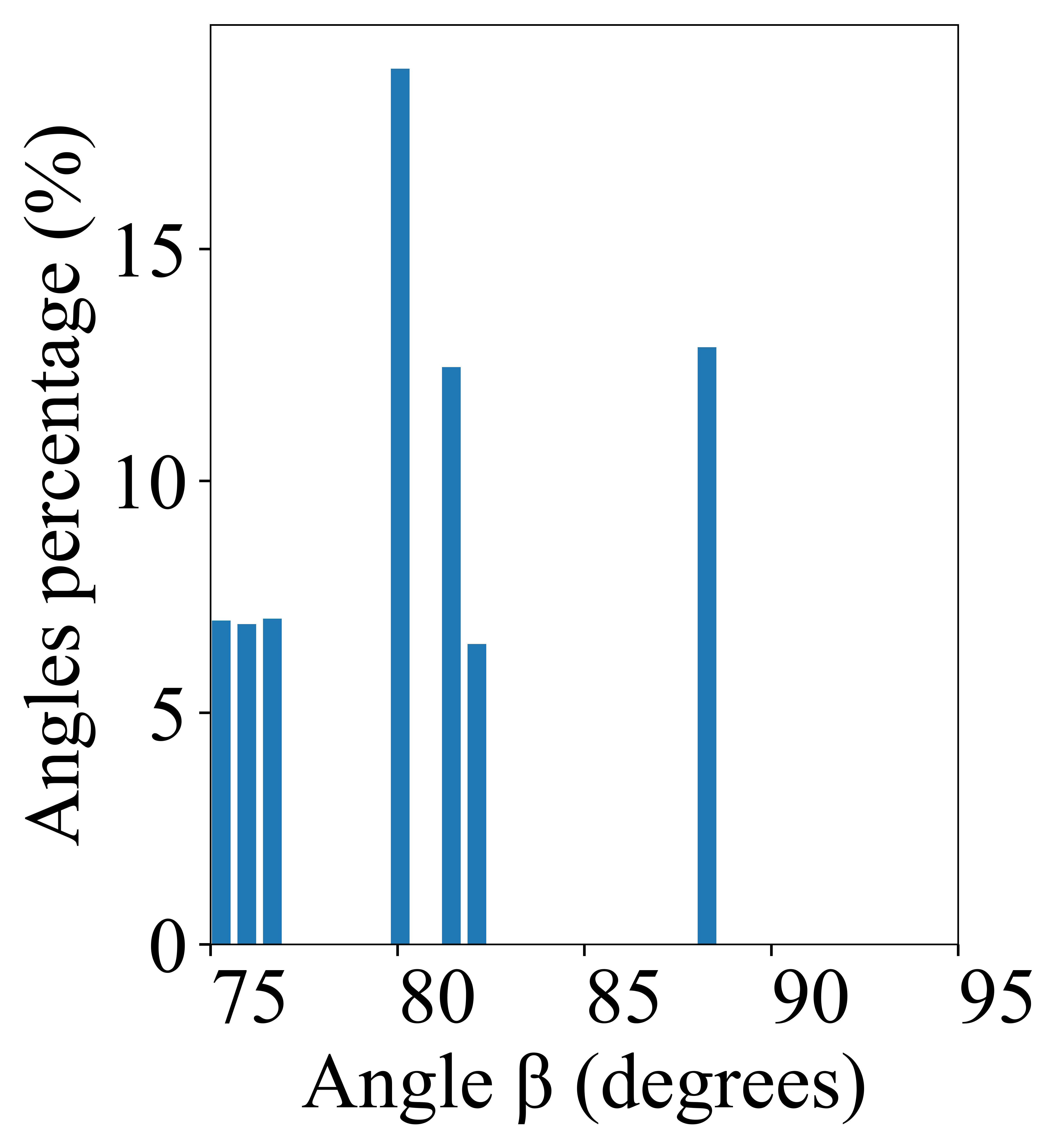}}
        (c)
	\end{minipage}

	\caption{Distributions of the angle $\beta$ between the planes and interplanar bonds in the Dn$\theta$ lattices ((a) - DnAB, (b) - Dn21.8, (c) - Dn27.8).}
	\label{fig:bond-angle-ch}
\end{figure*}

The reason for the difference in phonon lifetimes comes from the degree of disorder in the Moir\'e lattices. The structural disorder can be characterized by the variation of bond lengths in the structures. The distributions of the carbon-carbon (C-C) bond lengths for the Dn$\theta$ with $\theta$ of 0$^\circ$, 21.8$^\circ$, and 27.8$^\circ$ are shown in Fig.~\ref{fig:bond-len-ch}. The distribution of the $\beta$ angles of bonds with respect to the normal to the layers surface is shown in Fig.~\ref{fig:bond-angle-ch}. According to this data, the structures with higher twist angles have broader distributions of bond lengths. A similar correlation between twist angle values and distribution widths is observed in the BNn$\theta$, as shown in Figs. S7 and S8 in the Supplementary Information.

Quantitatively, this broadening is reflected in the increase of the standard deviation of bond lengths with the increase of the twist angle. The values of the standard deviations of the B-N and C-C bond lengths in the BNn$\theta$ and Dn$\theta$ structures are given in Table~\ref{table:bond_std}. One can note from Fig. \ref{fig:bond-angle-ch} that all interlayer C-C bonds in the 0$^\circ$ structures are perpendicular to the surface, while for the twisted lattices these angles between the planes and interplanar bonds start deviating from 90$^\circ$. The similar situation occurs in BNn$\theta$, as shown in Fig. S3 in the Supplementary information. This fact also proves that the degree of disorder grows with the increase of the twist angle in the considered Moir\'e lattices.

\begin{table}[h]
\caption{Standard deviations of B-N and C-C bond lengths in the BNn$\theta$ and Dn$\theta$ structures, respectively.}
\centering
\begin{tabular}{|c c c|}
\hline
  Twist angle $\theta$ & $\sigma(L_{B-N})$  & $\sigma(L_{C-C})$  \\ 
  \hline
  0$^\circ$ & 0.023 & 0.010 \\
  \hline
  21.8$^\circ$ & 0.028 & 0.025  \\
  \hline
  27.8$^\circ$ & 0.031 & 0.034  \\
  \hline
\end{tabular}
\label{table:bond_std}
\end{table}

The growing degree of disorder causes the increase of phonon scattering. This leads to the decrease of the phonon lifetimes and, as a consequence, should lead to lower LTC values (at fixed temperature).

\begin{figure}[h]
	\centering
	\begin{minipage}[h]{0.8\linewidth}
		\centering{\includegraphics[trim={0cm 0cm 0cm 0cm},clip, width=1\linewidth]{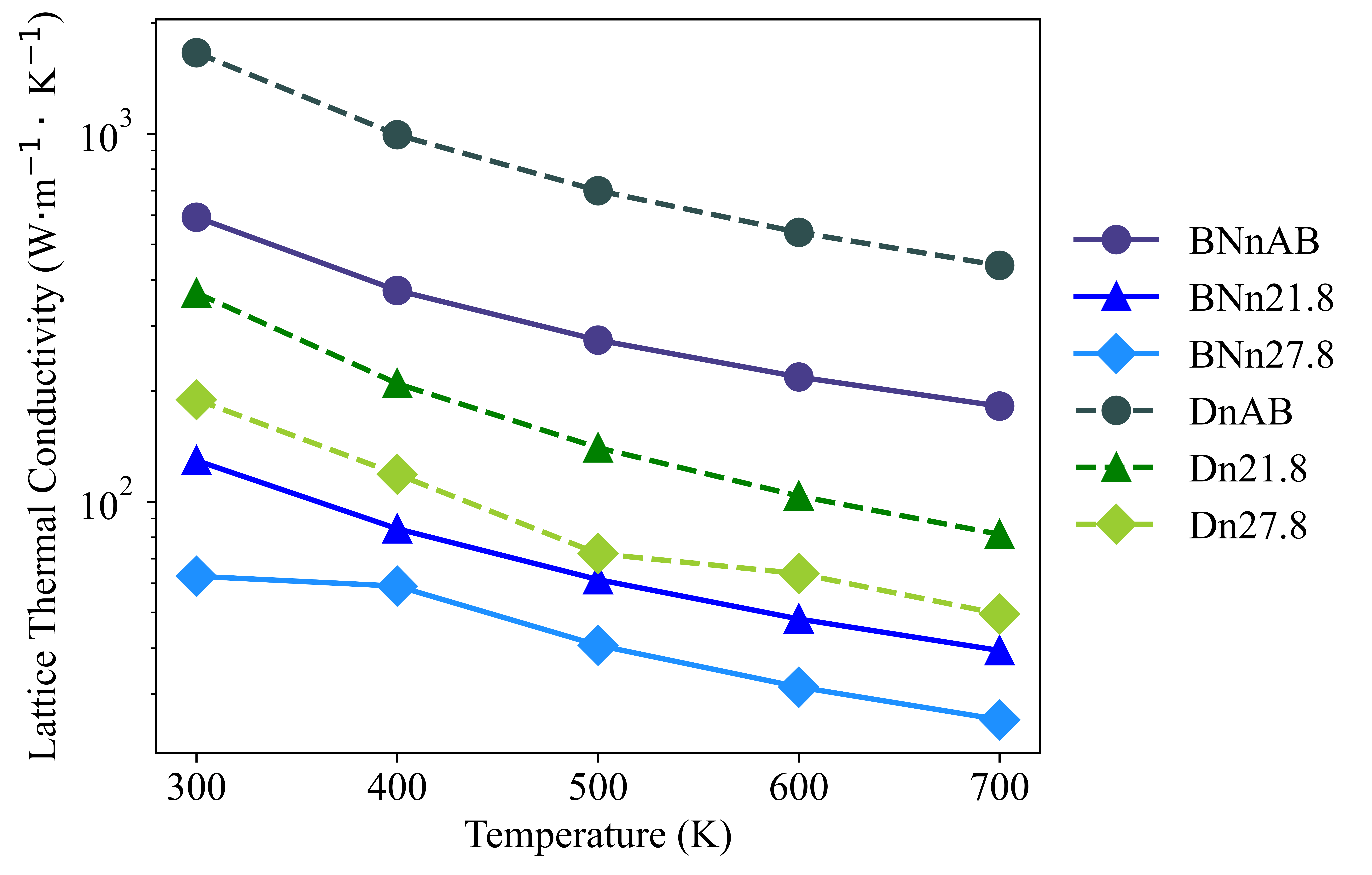}}

	\end{minipage}
	\caption{Temperature dependence of the lattice thermal conductivity in BNn$\theta$ and Dn$\theta$ structures. The converged grids used for solving phonon Boltzmann transport equation were  48$\times$48$\times$1,  24$\times$24$\times$1 and  14$\times$14$\times$1 for the Moir\'e lattices with the $\theta$ values of 0$^\circ$, 21.8$^\circ$ and 27.8$^\circ$, respectively. The convergence tests are described in the Supplementary information (see Fig. S1).}
	\label{fig:ltc_bnh}
\end{figure}

The results of LTC calculations for the BNn$\theta$ and Dn$\theta$ structures with different twist angles (0$^\circ$, 21.8$^\circ$, 27.8$^\circ$) are shown in Fig.~\ref{fig:ltc_bnh}.  

The dependence of calculated LTC values on the twist angle in Moir\'e lattices clearly shows the significant decrease of LTC with the twist angle growth. For example, the in-plane LTC of both Dn21.8 and BNn21.8 structures is reduced by a factor of 4.5 compared to the corresponding untwisted structures at 300 K. This reduction becomes more pronounced at the twist angle of 27.8$^\circ$, where the in-plane LTC at 300 K is approximately 9 times lower than that of the untwisted lattices. To sum up, our results demonstrate that increasing the twist angle in the Moir\'e lattices induces a greater degree of structural disorder. This increased disorder reduces phonon lifetimes, leading to the decrease in LTC. 

In addition, the decrease of LTC with increasing twist angle in hydrogenated graphene bilayers aligns with the findings of Chowdhury et al. \cite{kvashnin_diamanes} for Dn$\theta$ structures. However, the LTC values obtained in the present work for the Dn$\theta$ strucutres differ by 7-90\% from the values reported in \cite{kvashnin_diamanes}. This discrepancy arises because in the work of Chowdhury et al. the interatomic interactions were truncated at the fifth-nearest neighbor within the LTC calculation framework, whereas our BTE-based approach included all interatomic interactions. This difference underscores the importance of considering long-range interatomic interactions for accurate LTC predictions in these systems.

The BTE-based approach helped us to move from the LTC comparison to the comparison of several phonon properties (phonon group velocities, heat capacities and phonon lifetimes) and find which property change mostly contributes to the LTC decrease. However, while this method is useful for comparing LTC values for similar structures and finding the key reasons for LTC changes, the accuracy of BTE-based LTC calculations limited by three-phonon scattering is not guaranteed. For checking the applicability of this approach, the temperature dependence of the LTC was fitted by the power law relationship ($\kappa \sim T^\alpha$) for each Moir\'e lattice. The $\alpha$ values obtained for the Dn$\theta$ and BNn$\theta$ Moir\'e lattices are given in Table~\ref{table:alpha}.

\begin{table}[h]
\caption{Comparison of $\alpha$ values obtained for Dn$\theta$ and BNn$\theta$ structures.}
\centering
\begin{tabular}{|c c c|}

\hline
  Twist angle $\theta$ & $\alpha_{C-H}$  & $\alpha_{BN-H}$  \\ 
  \hline
  0$^\circ$ & -1.64 & -1.45 \\
  \hline
  21.8$^\circ$ & -1.81 & -1.43  \\
  \hline
  27.8$^\circ$ & -1.62 & -1.53  \\
  \hline
\end{tabular}

\label{table:alpha}
\end{table}

In the materials where three-phonon scattering is a dominant phonon scattering mechanism (i.e. BTE-based approach implemented in Phono3py is definitely valid), the values of the fitting parameter $\alpha$ are close to -1 \cite{about_alpha_fit}. However, according to the Tab. \ref{table:alpha}, the $\alpha$ values in the Dn$\theta$ and BNn$\theta$ significantly deviate from -1. This finding suggests that the BTE-based approach with only three-phonon processes is not sufficient to describe the thermal transport in the studied Moir\'e structures. For this reason, we moved to Green-Kubo method of LTC calculations that allows capturing higher-order scattering processes.

\begin{figure*}[h]
	\begin{minipage}[h]{0.5\linewidth}
		\centering{\includegraphics[trim={0cm 0cm 0cm 0cm},clip, width=1\linewidth]{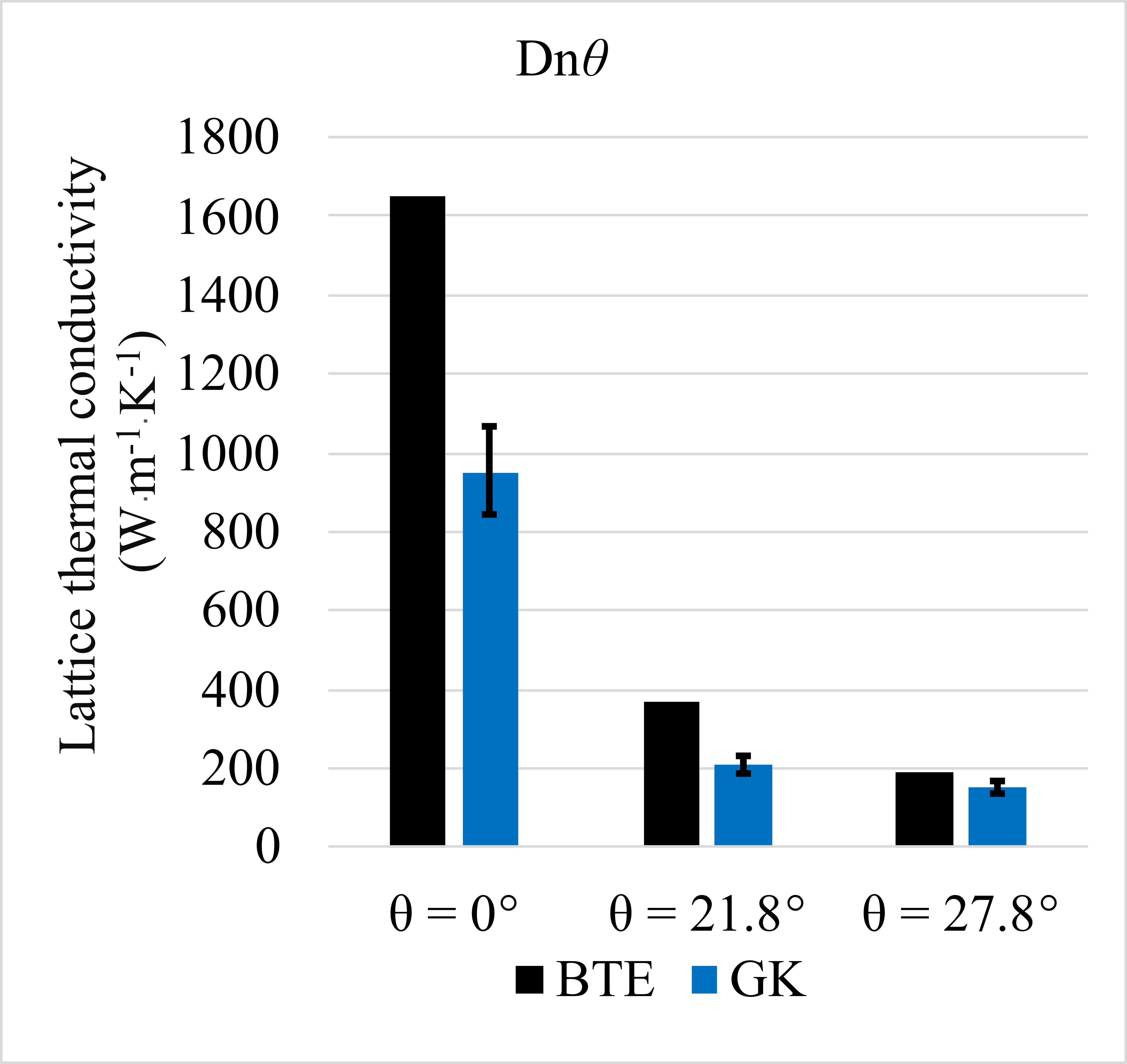}}
        (a)
	\end{minipage}%
 	\begin{minipage}[h]{0.5\linewidth}
		\centering{\includegraphics[trim={0cm 0cm 0cm 0cm},clip, width=1\linewidth]{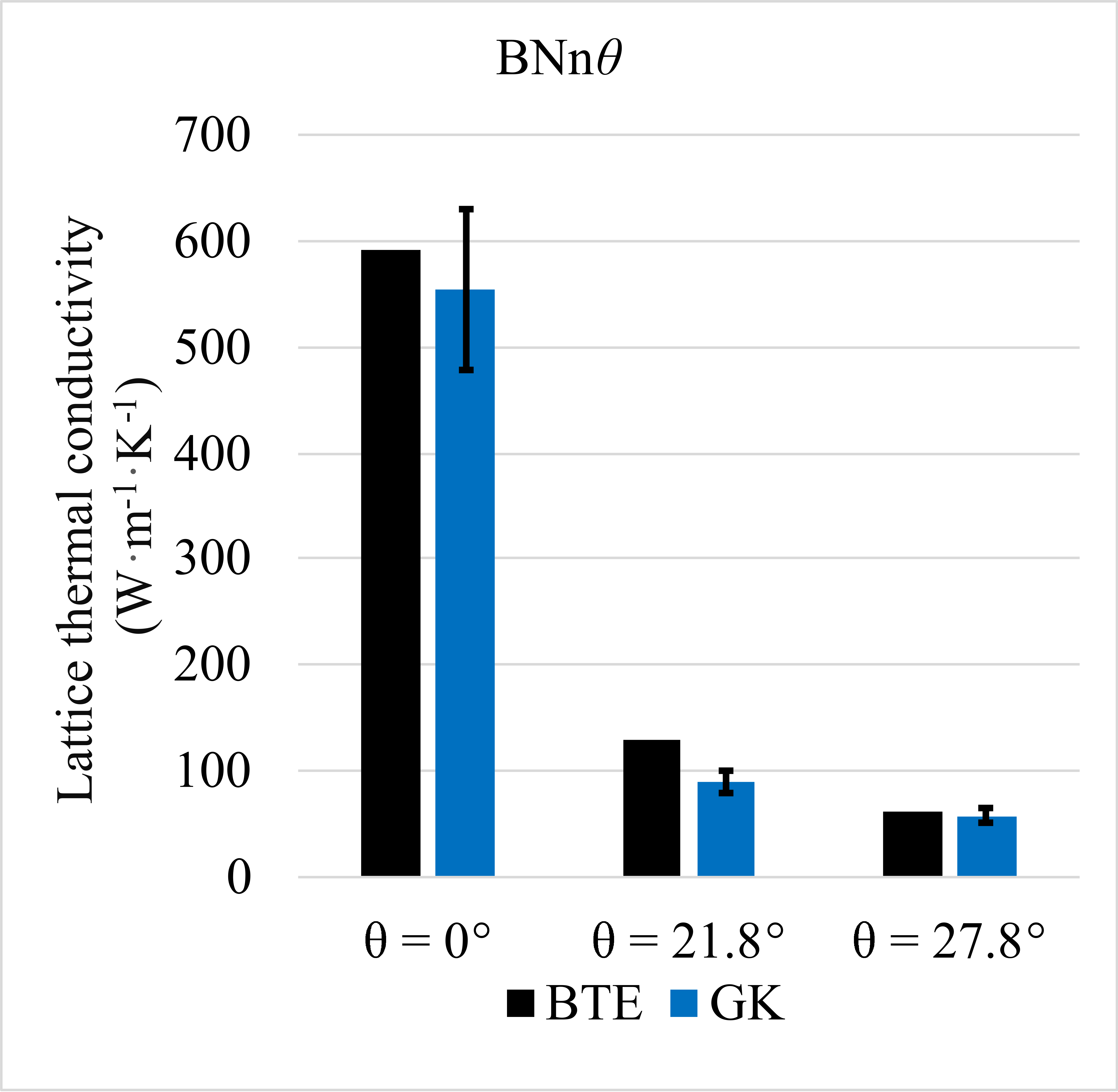}}
        (b)
	\end{minipage}

	\caption{Comparison of the LTC values obtained with BTE-based (BTE) approach and Green-Kubo method (GK) at T = 300 K.}
	\label{fig:GK_vs_BTE}
\end{figure*}

The results of LTC calculations obtained by means of the Green-Kubo method are presented in Fig. \ref{fig:GK_vs_BTE} and Tables \ref{table:ch_bte-gk}, \ref{table:bnh_bte-gk}. The LTC values differ significantly from those obtained with the BTE-based approach for all considered Dn$\theta$ and the BNn21.8. Therefore, the PES of these Moir\'e lattices possesses strong anharmonicity (even at T = 300 K), and the consideration of the fourth- and higher-order force constants for such materials is crucial for obtaining correct LTC values.

\begin{table}[h]
\caption{Comparison of the LTC values obtained with BTE-based (BTE) approach and Green-Kubo method (GK) at T = 300 K for the Dn$\theta$ structures.}
\centering
\begin{tabular}{|c c c c|}
\hline
  Twist angle $\theta$ & $\kappa_L^{GK}$ (W$\cdot$m$^{-1}\cdot$ K$^{-1}$) & $\kappa_L^{BTE}$ (W$\cdot$m$^{-1}\cdot$ K$^{-1}$) & $\Delta\kappa_L (\%)$ \\ 
  \hline
  0 & 951 $\pm$ 115 & 1653 & 41.1 \\
  \hline
  21.8 & 210 $\pm$ 21& 369 & 43.1 \\
  \hline
  27.8 & 153 $\pm$ 15& 189 & 19.0 \\
  \hline
\end{tabular}
\label{table:ch_bte-gk}
\end{table}

\begin{table}[h]
\caption{Comparison of the LTC values obtained with BTE-based (BTE) approach and Green-Kubo method (GK) at T = 300 K for the BNn$\theta$ structures.}
\centering
\begin{tabular}{|c c c c|}
\hline
  Twist angle $\theta$ & $\kappa_L^{GK}$ (W$\cdot$m$^{-1}\cdot$ K$^{-1}$) & $\kappa_L^{BTE}$ (W$\cdot$m$^{-1}\cdot$ K$^{-1}$) & $\Delta\kappa_L (\%)$ \\ 
  \hline
  0$^\circ$ & 555 $\pm$ 76 & 592 & 6.3 \\
  \hline
  21.8$^\circ$ & 89 $\pm$ 11& 129 & 31.0 \\
  \hline
  27.8$^\circ$ & 58 $\pm$ 6& 62 & 6.5 \\
  \hline
\end{tabular}
\label{table:bnh_bte-gk}
\end{table}

\subsection{Band gap renormalization}

The preceding discussion was focused on the effect of the interlayer twisting on the phonon properties and LTC of Moir\'e lattices. However, atomic vibrations are responsible not only for thermal transport. They also have a profound impact on electronic properties of materials through electron-phonon interactions. Specifically, we have examined the effects of twist angle on band gap, a key parameter influencing the electrical transport in materials. The band gap values calculated using the PBE functional at T = 0 K varied from 1.66 eV to 3.15 eV for the BNn$\theta$ structures and from 3.96 eV to 4.30 eV for the Dn$\theta$ structures (see Table S4 in the Supplementary Information). The corresponding electronic band structures are shown in Figs. S9 and S10 in the Supplementary Information.

\begin{figure*}[h]
	\begin{minipage}[h]{0.5\linewidth}
		\centering{\includegraphics[trim={0cm 0cm 0cm 0cm},clip, width=1\linewidth]{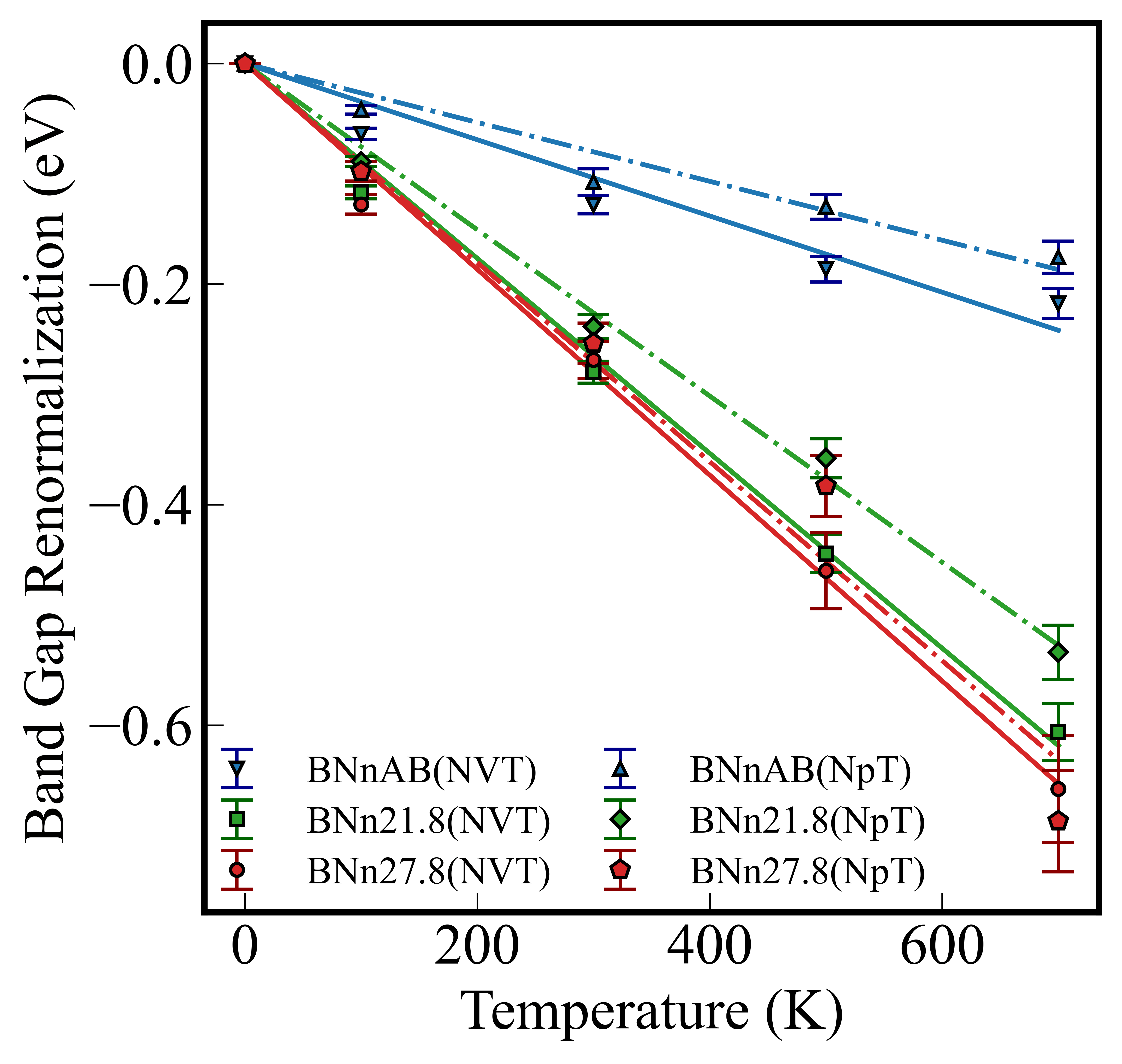}}
		(a)
	\end{minipage}%
	\begin{minipage}[h]{0.5\linewidth}
		\centering{\includegraphics[trim={0cm 0cm 0cm 0cm},clip, width=1\linewidth]{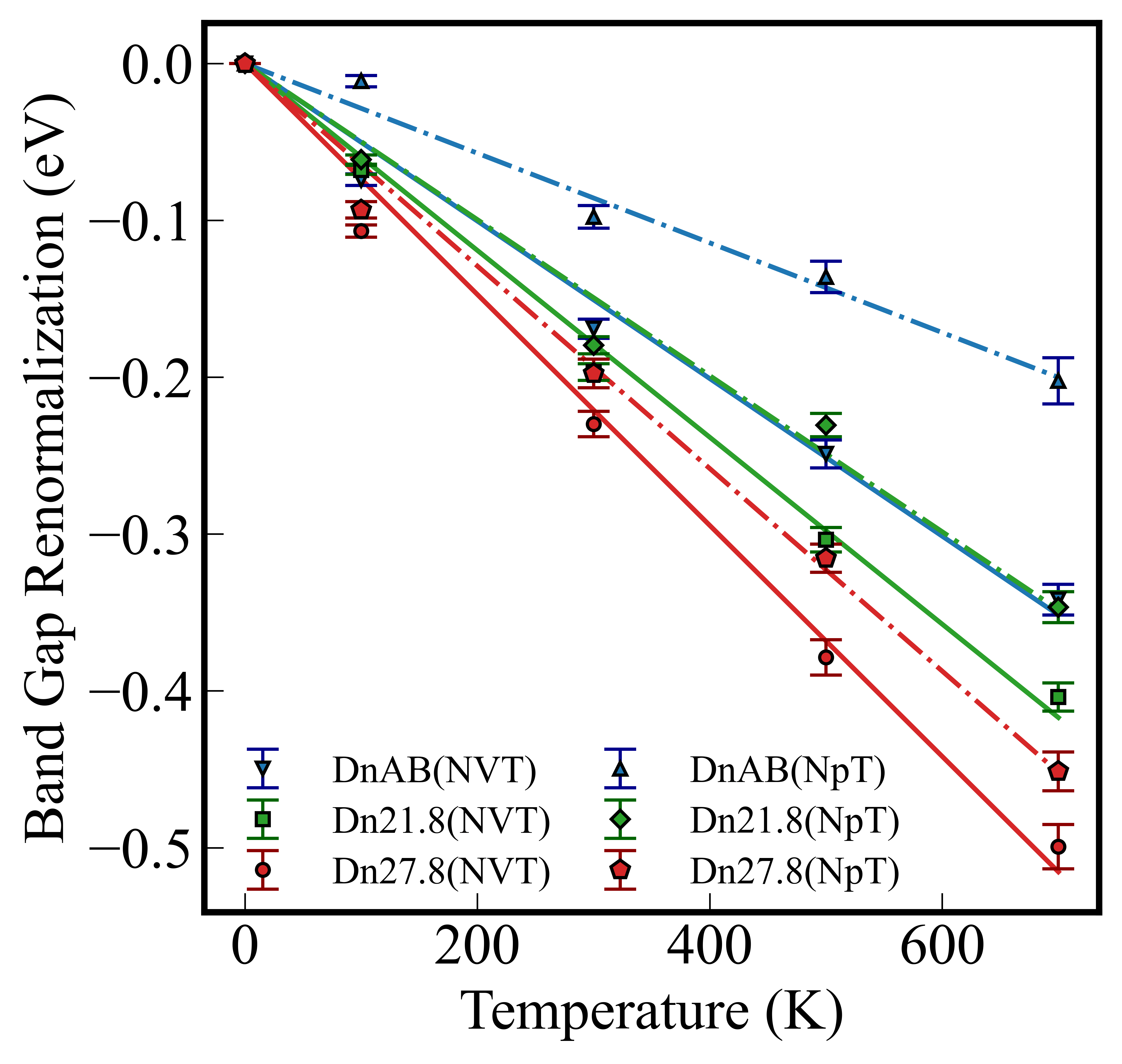}}
		(b)
	\end{minipage}
	
	\caption{Band gap renormalization in BNn$\theta$ (a) and Dn$\theta$ (b) Moir\'e lattices (BNn$\theta$/Dn$\theta$(NVT) - the distorted samples for calculating band gap values were obtained with classical NVT MD, fitted with line, BNn$\theta$/Dn$\theta$(NpT) - the distorted samples were generated by means of classical NpT MD, fitted with line.}
	\label{fig:bgr_lte}
\end{figure*}

\begin{figure*}[h]
	\begin{minipage}[h]{0.5\linewidth}
		\centering{\includegraphics[trim={0cm 0cm 0cm 0cm},clip, width=1\linewidth]{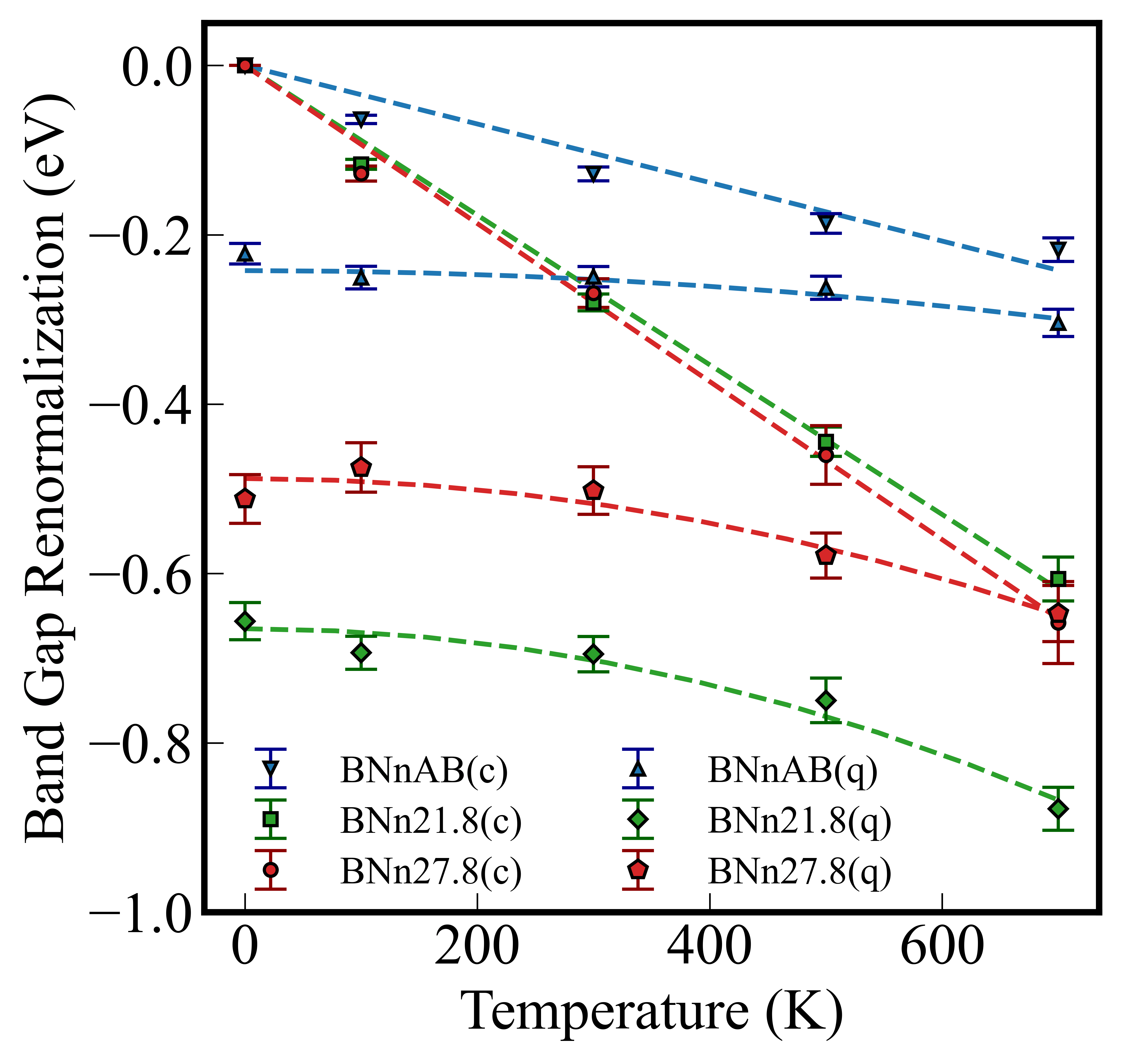}}
        (a)
	\end{minipage}%
 	\begin{minipage}[h]{0.5\linewidth}
		\centering{\includegraphics[trim={0cm 0cm 0cm 0cm},clip, width=1\linewidth]{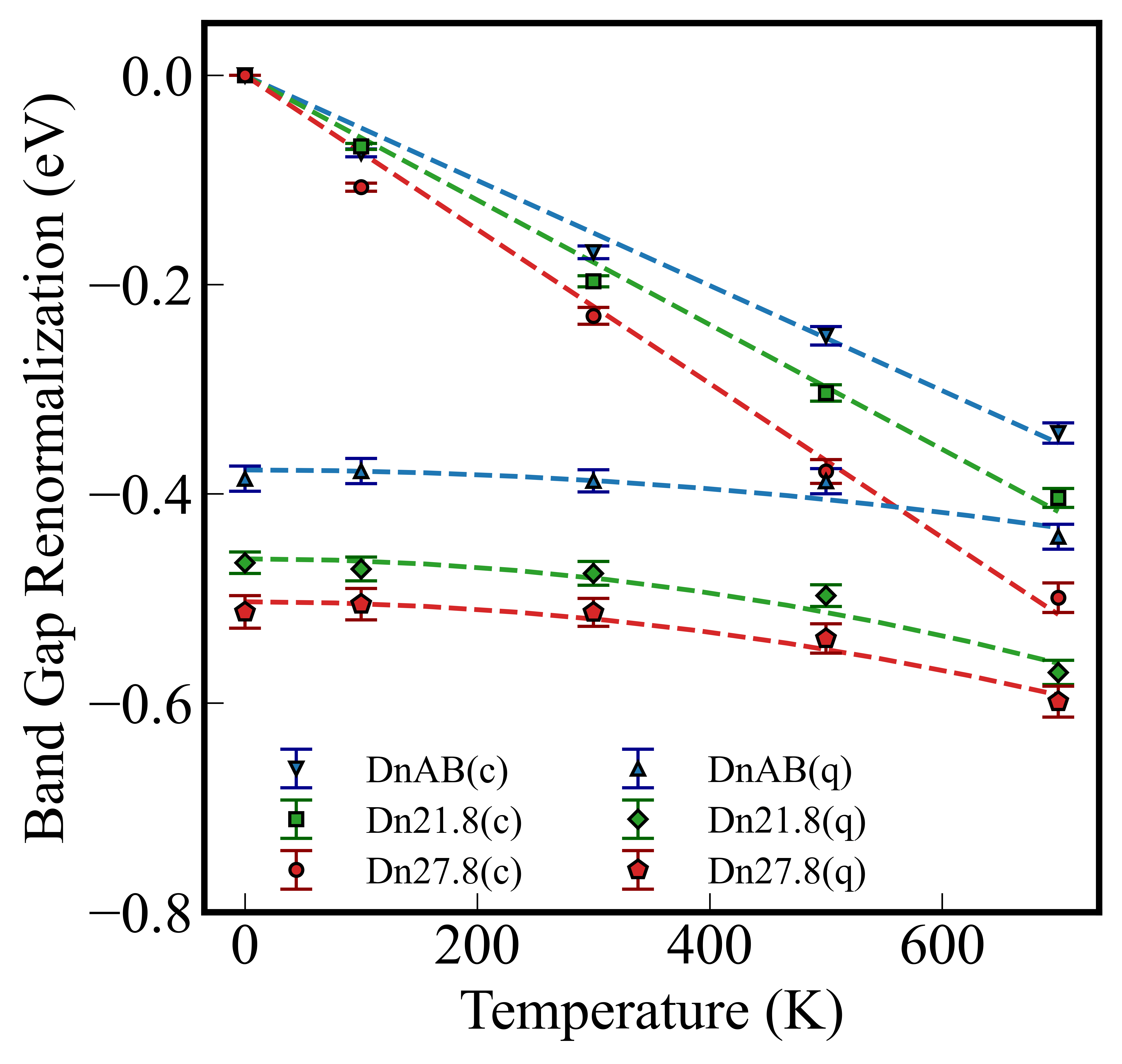}}
        (b)
	\end{minipage}

	\caption{Band gap renormalization in BNn$\theta$ (a) and Dn$\theta$ (b) Moir\'e lattices (BNn$\theta$/Dn$\theta$(c) - the distorted samples for calculating band gap values were obtained with classical NVT MD, fitted with line, BNn$\theta$/Dn$\theta$(q) - the distorted samples were generated by means of quantum harmonic sampling, as implemented in \cite{hiphive}), fitted with modified Varshni's expression \cite{varshni}.}
	\label{fig:bgr}
\end{figure*}

The results of the band gap renormalization calculations for the Dn$\theta$ and BNn$\theta$ Moir\'e lattices are presented in Figs. \ref{fig:bgr_lte} and \ref{fig:bgr}. In the case of classical nuclei, linear fitting of the BGR dependence on T was performed ($BGR(T) = a\cdot T$). For the quantum nuclei, modified form of Varshni's expression  was used for fitting $BGR(T)$ dependence:

\begin{equation}
    BGR(T) = ZPR - \frac{\alpha T^2}{T+\beta}.
\end{equation}

The details about Varshni's expression \cite{varshni} modification are given in the Supplementary Information.  As shown in Figs.\ref{fig:bgr_lte} and \ref{fig:bgr}, in both BNn$\theta$ and Dn$\theta$ cases, the band gap renormalization in the twisted structures is higher than in the untwisted ones. When the nuclei are considered as classical particles, higher values of the band gap renormalization correspond to the structures with higher twist angles. The analysis of mean-square displacements (MSD) of atoms in the structures with different twist angles showed that the MSD values were close for these structures (see Table S3 in Supplementary Information). Thus, the positive correlation between the twist angle and band gap renormalization was not caused by the difference in atomic displacements values. Therefore, the correlation can be attributed to the increase of disorder with the twist angle growth, similarly to the connection between the twist angle and the phonon lifetimes.
In addition, the effect of lattice thermal expansion on BGR was studied. According to Fig. \ref{fig:bgr_lte}, in both Dn$\theta$ and BNn$\theta$ structures LTE leads to smaller absolute values of BGR. This means that band gaps become wider due to LTE.

In the case of quantum nuclei, the values of ZPR in all Moir\'e lattices are high. These values are comparable with the ZPR in BN, diamond, and other materials with high ZPR, according to the work \cite{gonze_bgr_database}. These high values are caused by high frequencies of atomic vibrations typical for the materials consisting of light atoms (especially containing hydrogen). Passivation of the Moir\'e lattices surfaces with hydrogen atoms causes the presence of high-frequency phonon modes related to B-H, N-H and C-H bonds. These high-frequency vibrations have small impact on LTC (as discussed above), however, they significantly influence ZPR. 

In addition, for Dn$\theta$ the structures with bigger twist angles have higher ZPR values. However, in the case of BNn$\theta$ ZPR grows when we move from zero twist angle to 27.8$^\circ$ and then to 21.8$^\circ$. Similar trends are observed in the differences between BGR values obtained in classical and quantum cases for non-zero temperatures (see Fig. \ref{fig:bgr}). Looking at the data given in Table \ref{tab:zpr-compare}, one can notice that higher ZPR values are possessed by the structures with higher maximal vibrational frequencies (the phonon frequencies given in Table \ref{tab:zpr-compare} were obtained using MTP). However, if we look at the phonon band structures caluclated with DFT (Fig. \ref{fig:phonons_21_27}) we see that for BNn$\theta$ the maximal phonon frequencies grow with the twist angle increase. Thus, if the maximal frequencies and ZPR values are positively correlated, generation of rattled structures using the second-order force constants obtained from DFT forces followed by BGR calculations should lead to higher ZPR in BNn27.8 than in BNn21.8. In order to test this hypothesis, such calculations were performed. The resulting ZRP values were $0.68\pm0.02$ and $0.51\pm0.03$ for the BNn21.8 and BNn27.8 diamanes, respectively. Therefore, among the BNn$\theta$ structures BNn21.8 has the highest ZPR value, and it is not connected with its maximal phonon frequency. Due to this fact more detailed further studies of electron-phonon coupling in the Moir\'e diamanes are needed for explaining the influence of interlayer twisting on ZPR values.

\begin{table}[t]
\centering
\caption{Values of zero-point renormalization in Dn$\theta$, BNn$\theta$ Moir\'e lattices and the materials with comparable ZPR values studied in the work \cite{gonze_bgr_database}.}

\label{tab:zpr-compare}
\begin{tabular}{|lll|}
\hline
\multicolumn{3}{|c|}{Moir\'e lattices}                                          \\ \hline
\multicolumn{1}{|l|}{Material} & \multicolumn{1}{l|}{ZPR (eV)} & Maximal phonon frequency (THz) \\ \hline
\multicolumn{1}{|l|}{DnAB}   & \multicolumn{1}{l|}{0.39}  & 86.6 \\ \hline
\multicolumn{1}{|l|}{Dn21.8}  & \multicolumn{1}{l|}{0.47}  & 86.9 \\ \hline
\multicolumn{1}{|l|}{Dn27.8}  & \multicolumn{1}{l|}{0.51}  & 87.6 \\ \hline
\multicolumn{1}{|l|}{BNnAB}  & \multicolumn{1}{l|}{0.22}  & 98.9 \\ \hline
\multicolumn{1}{|l|}{BNn21.8} & \multicolumn{1}{l|}{0.66}  & 107.9 \\ \hline
\multicolumn{1}{|l|}{BNn27.8} & \multicolumn{1}{l|}{0.51}  & 103.7 \\ \hline
\multicolumn{3}{|c|}{Materials described in the work \cite{gonze_bgr_database}}                                           \\ \hline
\multicolumn{1}{|l|}{Material}                  & \multicolumn{2}{l|}{ZPR (eV)} \\ \hline
\multicolumn{1}{|l|}{C (diamond)}              & \multicolumn{2}{l|}{0.33}     \\ \hline
\multicolumn{1}{|l|}{BN (zinc blende)}         & \multicolumn{2}{l|}{0.41}     \\ \hline
\multicolumn{1}{|l|}{SiC (zinc blende)}         & \multicolumn{2}{l|}{0.18}     \\ \hline
\end{tabular}
\end{table}

\section{Conclusions}

The effect of interlayer twisting on lattice dynamics, thermal transport and electron-phonon coupling in diamane-like BNn$\theta$ and Dn$\theta$ Moir\'e lattices was systematically investigated. The calculations were done by means of the density functional theory and machine learning moment tensor interatomic potentials. The comparison of LTC values obtained with GK and BTE methods showed that anharmonic effects related to  four-phonon and other higher-order phonon-phonon interactions significantly affect LTC in the three Dn$\theta$ and the BNn21.8 diamanes. By comparing the LTC for the structures with different twist angles, we have found that the increase of twist angle leads to the growth of the structure disorder in both Dn$\theta$ and BNn$\theta$ Moir\'e lattices. This leads to the decrease of the phonon lifetimes and, as a consequence, a decrease of the LTC. Another consequence of the structure disorder growth is the increase of band gap renormalization caused by classical nuclei motion. Significant ZPR values observed in the Moir\'e lattices are caused by the presence of light hydrogen atoms on the lattices surface. These findings highlight the potential of twist angle and composition engineering as a powerful tool for simultaneous control of thermal transport and electronic properties in Moir\'e lattices, paving the way for the development of novel materials with tailored properties for thermoelectric, optoelectronic, and microelectronic devices.

{\it Acknowledgments.}
R.A., N.R., M.P., and A.S. acknowledge funding from the Russian Science Foundation (Project No. 23-13-00332). V.D. and L.C. acknowledge funding from the Russian Ministry of Science and Higher Education (Project No. 125020401357-4).


\end{document}